\documentclass{aa}  
\usepackage{natbib}
\usepackage{braket}
\usepackage{graphicx}
\usepackage{booktabs}
\usepackage{setspace}
\usepackage[colorlinks=true,linkcolor=blue,citecolor=blue, urlcolor=blue]{hyperref}
\usepackage{txfonts}
\newcommand\HI{H {\small{I}}}

\begin{document}

   \title{Semi-empirical constraints on the H{\Large I} mass function of star-forming galaxies and $\Omega_{\rm HI}$ at $z\sim 0.37$ from interferometric surveys}

   \titlerunning{Constraints on $\Omega_{\rm HI}$ and the HI mass function at $z\sim 0.37$}

   \author{F. Sinigaglia      \inst{1,2}\fnmsep\thanks{\email{francesco.sinigaglia@unige.ch}}
    \and
          A. Bianchetti\inst{3,4}
          \and
          G. Rodighiero\inst{3,4}
          \and
          L. Mayer\inst{2}
          \and
          M. Dessauges-Zavadsky\inst{1}
          \and
          \\
          E. Elson\inst{5}
          \and
          M. Vaccari\inst{6,7,8}
          \and
          M. J. Jarvis\inst{9,10}
          }

    \authorrunning{F. Sinigaglia et al.}

   \institute{Département d’Astronomie, Université de Genève, Chemin Pegasi 51, CH-1290 Versoix, Switzerland
        \and Department of Astrophysics, University of Zurich, Winterthurerstrasse 190, CH-8057 Z\"urich, Switzerland
        \and
        Department of Physics and Astronomy, Università degli Studi di Padova, Vicolo dell’Osservatorio 3, I-35122, Padova, Italy
        \and
        INAF - Osservatorio Astronomico di Padova, Vicolo dell’Osservatorio 5, I-35122, Padova, Italy
        \and 
        Department of Physics and Astronomy, University of the Western Cape, Robert Sobukwe Rd, 7535 Bellville, Cape Town, South Africa
        \and
        Inter-university Institute for Data Intensive Astronomy, Department of Astronomy, University of Cape Town, 7701 Rondebosch, Cape Town, South Africa
        \and
        Inter-university Institute for Data Intensive Astronomy, Department of Physics and Astronomy, University of the Western Cape, 7535 Bellville, Cape Town, South Africa
        \and
        INAF - Istituto di Radioastronomia, via Gobetti 101, 40129 Bologna, Italy
        \and
        Astrophysics, University of Oxford, Denys Wilkinson Building, Keble Road, Oxford OX1 3RH, UK
        \and
        Department of Physics and Astronomy, University of the Western Cape, Cape Town 7535, South Africa
             }

   \date{Received \today; accepted XYZ}

 
  \abstract
    {The H {\scriptsize I} mass function is a crucial tool to understand the evolution of the H {\scriptsize I} content in galaxies over cosmic times, and hence, to constrain both the baryon cycle in galaxy evolution and the reionization history of the Universe.}
   {We aim to derive semi-empirical constraints at $z\sim 0.37$ by combining literature results on the stellar mass function from optical surveys with recent findings on the $M_{\rm HI}-M_\star$ scaling relation derived via spectral stacking analysis applied to 21-cm line interferometric data from the MIGHTEE and CHILES surveys, conducted with the MeerKAT and VLA radio telescopes, respectively.}
   {We draw synthetic stellar mass samples directly from the publicly-available results underlying the analysis of the COSMOS2020 galaxy photometric sample. Afterwards, we convert $M_\star$ into $M_{\rm HI}$ using analytical fitting functions to the data points from H {\scriptsize I} stacking. We then fit a Schechter function to the median HIMF from all the samples via Markov Chains Monte Carlo. We finally derive the posterior distribution for $\Omega_{\rm HI}$ by integrating the models for the HIMF built from the posteriors samples of the Schechter parameters.}
   {We find a deviation of the HIMF at $z\sim 0.37$ from the results at $z\sim 0$ from the ALFALFA survey and at $z\sim 1$ from uGMRT data. Our results for $\Omega_{\rm HI}$ are in broad agreement with other literature results, and follow the overall trend on $\Omega_{\rm HI}$ as a function of redshift. The derived value $\Omega_{\rm HI}=\left(7.02^{+0.59}_{-0.52}\right)\times10^{-4}$ at $z\sim 0.37$ from the combined analysis deviates at $\sim 2.9\sigma$ from the ALFALFA result at $z\sim 0$.}
   {Our findings about the HIMF and $\Omega_{\rm HI}$ derived from deep, state-of-the-art interferometric surveys differ from previous literature results at $z\sim0$ and $z\sim1$, although we are unable to confirm at this stage whether these differences are due to cosmic evolution consistent with a smooth transition of the \HI{} content of galaxies over the last 8 Gyr or due to selection biases and systematics.}


   \maketitle 
%
\section{Introduction}

The neutral atomic hydrogen (hereafter \HI{}) plays a key role in astrophysics and cosmology. In fact, the distribution of \HI{} in the Universe is driven by the complex interplay of astrophysical and cosmological effects. The abundance of \HI{} in the large-scale structure is regulated by the nonlinear formation and evolution of the cosmic web \citep[e.g.,][]{Martizzi2019,Galarraga2021,Sinigaglia2021,Sinigaglia2024}. Once galaxies form within dark matter haloes thanks to the conversion of the intra-halo \HI{} to molecular hydrogen and to gravitational collapse \citep[e.g.,][]{Somerville2015,Vogelsberger2020,Primack2024}, the \HI{} constitutes the main raw fuel for star formation and is one of the major ingredients contributing to the baryon cycle and in regulating the galaxy mass assembly \citep[e.g.,][]{Blitz2006,Bigiel2008,Sternberg2014}. For these reasons, constraining the \HI{} cosmic density parameter $\Omega_{\rm HI}$ and the \HI{} mass function (hereafter HIMF) over cosmic time is of paramount importance to try answering profound questions about the Universe, such as probing cosmic reionization \citep[e.g.,][]{Koopman2015,DeBoer2017}, understanding the galaxy mass assembly and the time evolution of the star formation rate density \citep[e.g.,][]{Huang2012,Maddox2015,Catinella2018,Sinigaglia2022,Bianchetti2025}, and even using \HI{} to infer cosmological information  such as e.g. unveiling the nature of dark matter \citep[e.g.,][]{Pinetti2020,Bauer2021,Banares2023} and of dark energy \citep[e.g.,][]{Santos2015,Zhang2023,Berti2024}, as well as testing the $\Lambda$CDM model and other alternative theories of gravity with resolved galaxy rotation curves \citep[e.g.,][]{OBrien2018}.

Measuring $\Omega_{\rm HI}$ and the HIMF is, however, a challenging task. Due to the intrinsic faintness of the 21-cm line and the limited sensitivity of the majority of the existing radio telescopes, a direct measurement of the \HI{} emission has so far been possible only out to relatively low redshift. Current existing constraints on $\Omega_{\rm HI}$ and on the HIMF based on a statistically large sample of \HI{} galaxies are in fact limited to the nearby universe ($z\lesssim 0.1$) \citep[e.g.,][]{Zwaan2003,Martin2010,Jones2018,Ponomareva2023}. Among these, the most accurate measurement of the HIMF is the one presented by the full ALFALFA survey at $z\le 0.06$, employing approximately $31,500$ \HI{} sources \citep{Jones2018}. At $z>0.1$ only a few studies have reported the measurements of the HIMF and of $\Omega_{\rm HI}$ based on direct \HI{} detection, such as the one from the AUDS survey at $z=0.16$ \citep{Hoppman2015} and from the BUDHIES survey at $z= 0.2$ \citep{Gogate2022}. Nonetheless, the available number of sources is still limited to a few tens/hundreds, implying a considerable uncertainty, especially on the high-mass end of the HIMF. In addition, BUDHIES targeted dense fields, which may not be representative of the global galaxy population. 
Alternative approaches to estimating $\Omega_{\rm HI}$ involves either measuring the average \HI{} mass in galaxies \citep[e.g.][]{Lah2007,Delhaize2013,Rhee2016,Rhee2018,Hu2019}, using the cross correlation between optical galaxies and \HI{} intensity mapping \citep{Chang2010,Masui2013,Wolz2022,Cunnington2023}, or relying on the \HI{} absorption features observed in Damped Ly$\alpha$ systems \citep[e.g.,][]{Rao2006,Noterdaeme2012,Chrighton2015,Rao2017}. These methods, however, do not yield a measurement of the HIMF, but only an estimate of the total \HI{} cosmic density.

Recently, a novel semi-empirical method --- only possible thanks to the improved sensitivity of state-of-the art radio telescopes such the MeerKAT and the upgraded Giant Metrewave Radio Telescope (uGMRT) --- has emerged as an indirect but cheap alternative to the direct \HI{} detection to constrain $\Omega_{\rm HI}$ and the HIMF at intermediate and high redshifts. This method consists of exploiting robust measurements from the literature of the galaxy stellar mass or luminosity function across redshift and to convert it to \HI{} via a scaling relation, measured with a \HI{} stacking approach. \cite{Bera2022} and \cite{Chowdhury2024} pioneered this technique and used it to measure the HIMF at $z\sim 0.35$ and $z\sim 1$, respectively. In particular, therein the authors exploit the $B$-band magnitude $M_B$ luminosity function and use the $M_{\rm HI}-M_{\rm B}$ scaling relation for star-forming galaxies that they measure using a 21-cm line spectral stacking approach from uGMRT observations to convert from $M_B$ to $M_{\rm HI}$ and obtain the HIMF. 

Eventually, the \HI{} intensity mapping technique is also capable of providing (indirect) constraints on $\Omega_{\rm HI}$ and the HIMF. In this sense, \cite{Paul2023} reported the first measurement of the HIMF from a \HI{} intensity mapping experiment. Specifically, therein the authors presented the first measurement of the \HI{} intensity mapping power spectrum at $z\sim 0.32$ and used  a halo model approach to constrain the HIMF, following \cite{Chen2021}.

In this work, we follow a methodology similar in spirit to the one by \cite{Bera2022} and \cite{Chowdhury2024}. In particular, we combine the most recent measurements of the stellar mass function of star-forming galaxies in the COSMOS field at $0.2<z<0.5$ by \cite{Weaver2023} and of the $M_{\rm HI}-M_\star$ scaling relation of star-forming galaxies at $0.22<z<0.49$ (median redshift $z\sim 0.37$) by \cite{Bianchetti2025} from the \HI{} stacking analysis of two state-of-the-art deep \HI{} interferometric surveys MIGHTEE \citep{Jarvis2016} and CHILES \citep{Fernandez2013}, both covering the COSMOS field \citep{Scoville2007}. With these tools, we derive novel indirect semi-empirical constraints on the HIMF of star-forming galaxies and the cosmic \HI{} density parameters $\Omega_{\rm HI}$ at $z\sim 0.37 $.

The paper is organized as follows. Sect. \ref{sec:data} introduces the optical and \HI{} data used in this work. Sect. \ref{sec:refsim} summarizes the main features of the cosmological hydrodynamic simulations we use to compare our results with. We present our methodology in Sect. \ref{sec:methods}. Sect. \ref{sec:results} presents the analysis, the results of our predictions and a discussion of them. We conclude in Sect. \ref{sec:conclusions}.

Throughout this work, we adopt the following cosmological parameters: $\Omega_m=0.3$, $\Omega_\Lambda=0.7$, $H_0=70.0~{\rm km}~{\rm s}^{-1}~{\rm Mpc}^{-1}$. When relevant, we have rescaled the results from other literature results to the match the cosmological parameters adopted herein.


\section{Data from optical and \HI{} surveys in the COSMOS field}\label{sec:data}

In this section, we summarize the main features of the data from optical surveys included in the COSMOS2020 catalog \citep{Weaver2022}, from the MIGHTEE and the CHILES 21-cm line interferometric surveys, as well as the main stacking results from \cite{Bianchetti2025}. We refer to \cite{Bianchetti2025}, for a more detailed description of all the items described in what follows.

\subsection{The COSMOS2020 sample and the stellar mass function}\label{sec:cosmos2020}

The optical catalog underlying the stacking analysis performed by \cite{Bianchetti2025} was built through a cross-match between the COSMOS2020-CLASSIC\footnote{Release 1, v2.2, March 2023} photometric catalog \citep{Weaver2022} and a spectroscopic catalog assembled  by merging three different catalogs covering the COSMOS field: the COSMOS spec-$z$ compilation \citep{Khostovan2025}, the DEVILS survey catalog \citep{Davies2018} and the DESI survey Early Data Release catalog \citep{DESI2023}. 
First, \cite{Bianchetti2025} selected only spectroscopic sources in the COSMOS field \citep{Scoville2007} and at $0.22<z<0.49$ (median redshift $z\sim 0.37$), i.e. within a subset of the volume probed by the MIGHTEE and CHILES data sets (see Sect. \ref{sec:mightee_chiles}). In particular, they performed a positional cross-match (with matching radius=$1$ arcsec) between the spectroscopic and the photometric catalog. The photometric catalog includes galaxy properties obtained through spectral energy density (SED) fitting performed with the \texttt{LePhare} code \citep{Ilbert2006,Arnouts2011}. Therefore, after the cross-match, they retrieved from the photometric catalog all the aforementioned photometrically-derived parameters for each of the sources in the spectroscopic catalog. Star-forming galaxies were then selected using the in-built color-color selection in the $(NUV-r)/(r-J)$ rest-frame plane. Further quality cuts were then applied to ensure the highest possible reliability, including redshift quality cuts and photometric redshift outliers exclusion.  

\cite{Weaver2022} and \cite{Weaver2023} have shown that the COSMOS photometric sample is complete in stellar mass down to $\log_ {10}(M_{\star}/{\rm M}_{\odot}) \approx 8$ at $z < 0.5$. 

In this work, we also make use of the results on the stellar mass function in the COSMOS field as measured from the COSMOS2020 catalog \citep{Weaver2023}. In particular, \cite{Weaver2023} performed a detailed study of the stellar mass sample from  COSMOS2020, deriving the stellar mass functions for the total, star-forming, quiescent populations at $0.2<z<0.5$, as well as in several higher redshift ranges. The splitting of the galaxy population into star-forming and quiescent was performed according to the same color criterion mentioned above. These aspects, together with the fact that the underlying photometric catalog is the same, make the stellar mass function derived by \cite{Weaver2023} as the ideal candidate to be used in this work. \cite{Weaver2023} used a Markov Chain Monte Carlo method to fit a double Schechter profile to the resulting stellar mass function:
\begin{equation}
\label{eq:double_schechter}
\begin{split}
    \Phi ~ d\log_{10}&(M_\star/{\rm M}_{\odot}) = \log(10)\exp\left(-10^{\log_{10}(M_\star/{\rm M}_{\odot})-\log_{10}(\tilde{M}/{\rm M}_{\odot})}\right) \\
    & \times \biggl[  \tilde{\Phi}_1 \left(10^{\log_{10}(M_\star/{\rm M}_{\odot})-\log_{10}(\tilde{M}/{\rm M}_{\odot})}\right)^{\alpha_1 +1}\\
    & +  \tilde{\Phi}_2 \left(10^{\log_{10}(M_\star/{\rm M}_{\odot})-\log_{10}(\tilde{M/{\rm M}_{\odot}})}\right)^{\alpha_2 +1} \biggr]  ~d\log_{10}(M_\star/{\rm M}_{\odot}) \quad .
\end{split}
\end{equation}

For the star-forming mass-complete galaxy subsample at $0.2<z<0.5$, \cite{Weaver2023} found the following best-fitting parameters:
\begin{multline*}
    \{\log_{10}(\tilde{M}/{\rm M}_{\odot}),~ \alpha_1, ~\tilde{\Phi}_1\times 10^{-3},~ \alpha_2,~ \tilde{\Phi}_2\times 10^{-3}\}= \\
    = \{ 10.73^{+0.17}_{-0.15},~ -1.41^{+0.03}_{-0.04},~ 0.80^{+1.37}_{-0.56},~ -0.02^{+0.61}_{-0.79},~ 0.49^{+0.32}_{-0.33}\} \quad .
\end{multline*}

The best-fitting values quoted above consist of the median of the posteriors of the parameters. While the median stellar mass function can then be readily obtained by substituting those values for the parameters inside Eq. \ref{eq:double_schechter}, we would not be properly accounting for the parameters uncertainty and covariance. In Sect. \ref{sec:methods}, we will outline the Monte Carlo procedure used throughout this work, in which we need to obtain many samples for the stellar mass function varying the best-fitting parameters within their uncertainty, and hence, correctly accounting for their covariance. To do so, instead of resorting to the best-fitting median values of posteriors, we randomly draw samples directly from the Markov Chain posteriors by \cite{Weaver2023}, made publicly available\footnote{\url{https://zenodo.org/records/8377094}}. This allows us to correctly account for uncertainties on the stellar mass function in a Bayesian fashion within our framework.  

\subsection{The MIGHTEE and CHILES surveys}\label{sec:mightee_chiles}

MIGHTEE \citep{Jarvis2016} is a survey conducted with MeerKAT radio telescope. The MeerKAT radio interferometer \citep{Jonas2016} consists of 64 offset Gregorian dishes equipped with receivers in UHF–band ($580$ MHz $< \nu <$ $1015$ MHz), L–band (900 MHz $< \nu < $1670 MHz), and S–band (1750 MHz$< \nu <$ 3500 MHz), and serves as a precursor to the Square Kilometre Array (SKA). MIGHTEE is an L-band continuum, polarization, and spectral-line large survey conducted with MeerKAT, utilizing spectral and full Stokes mode observations. It covers four deep extragalactic fields (COSMOS, XMM-LSS, ECDFS, ELAIS-S1), chosen because of their extensive multi-wavelength coverage from previous and ongoing observations. For this paper, we use results derived for the MIGHTEE-\HI{} Early Science spectral line data from MIGHTEE \citep{Maddox2021} covering the COSMOS field with a single pointing within the redshift range of $0.22 < z < 0.49$, with an effective exposure time of approximately 23 hours. The MIGHTEE beam is approximately $17.2^{\prime\prime}\times 13.9^{\prime\prime}$ at $z\sim 0.37$ ($\sim 90 \times 73~ {\rm kpc}^2$).
The median \HI{} noise RMS of the cubes increases as the frequency decreases, ranging from 85 $\rm \mu Jy~beam^{-1}$ at $\nu \sim 1050~ {\rm MHz}$ to $135~\mu {\rm Jy}~{\rm beam}^{-1}$ at $\nu \sim 950~{\rm MHz}$ at a spectral resolution (channel width) of $209~{\rm kHz}$.

CHILES is a deep-field \HI{} survey, carried out with the VLA and imaging \HI{} over a contiguous redshift range of $0<z<0.49$ for the first time \citep{Fernandez2013}. The pointing was centered on RA(J2000) 10h01m24s and DEC(J2000) 2d21m00s. It samples a subregion of the COSMOS field; therefore, it provides an independent measurement of the same patch of sky as the MIGHTEE survey. The 25-meter diameter antennas of the VLA provide a field of view (FOV) of around $0.5$ deg at $1.4$ GHz. Due to the rotating VLA configurations, the CHILES observations were split into five observing epochs spaced approximately 15 months apart. In this paper, we use results from successfully processed data from all five of the observing epochs, amounting to approximately $800$ observation hours (approximately $600$ on-source hours). The VLA-B configuration was used, achieving a beam size of approximately $7.2^{\prime\prime}\times 6.4^{\prime\prime}$ at $z\sim 0.37$ ($\approx 38 \times 33$ ${\rm kpc^2}$). The datacube comprises 125 kHz wide spectral channels at $z\sim 0.37$, covering a wide range of frequencies ($950-1420~{\rm MHz}$). The RMS ranges between $30$ and $50~\mu {\rm Jy~beam^{-1}}$ in this range.

\subsection{Summary of stacking results from MIGHTEE and CHILES}\label{sec:ref_data}

A fundamental ingredient upon which this work is based consists of the stacking results from \cite{Bianchetti2025}. In particular, as anticipated, therein the authors perform a 21-cm line stacking analysis in different stellar mass bins for the star-forming galaxy population, with the aim of measuring the $M_{\rm HI}-M_\star$ scaling relation from the combination of the MIGHTEE and CHILES surveys. While a detailed summary of the results goes beyond the scope of this paper (we refer directly to \cite{Bianchetti2025} for it), we report in Table \ref{tab:stacking_results} the numerical values for the data points. In particular, we notice that for the combined stacking the larger statistics allowed the authors to split the sample into $4$ stellar mass bins, while for the MIGHTEE and CHILES separate stacking analyses, it was possible to split the galaxy population only into $3$ stellar mass bins to be able to detect the \HI{} signal in all of them. This aspect will become important later on in the paper when describing the fitting strategy for the $M_{\rm HI}-M_\star$ relation.

\begin{table*}[]
    \centering
    \begin{tabular}{lccc}
    \toprule
    $\log_{10}(M_\star/{\rm M}_{\odot})$ bin & Median $\log_{10}(M_\star/{\rm M}_{\odot})$ & $N_{\rm gal}$ & $\braket{\log_{10}(M_{\rm HI}/{\rm M}_{\odot})}$\\
    \midrule
    & {\bf Combined} & \\
    \midrule
    $8.0<\log_{10}(M_{\star}/{\rm M_{\odot}})<9.5$ & $9.10$ & $3507$ & $9.51\pm 0.05$\\
    $9.5<\log_{10}(M_{\star}/{\rm M_{\odot}})<9.8$ & $9.65$ & $909$ & $9.68\pm 0.06$\\
    $9.8<\log_{10}(M_{\star}/{\rm M_{\odot}})<10.5$ & $10.17$ & $1564$ & $9.94 \pm 0.02$\\
    $\log_{10}(M_{\star}/{\rm M_{\odot}})>10.5$ & $10.85$ & $636$ & $10.07\pm 0.04$\\
    \midrule
    & {\bf MIGHTEE} & \\
    \midrule
    $8.0<\log_{10}(M_{\star}/{\rm M_{\odot}})<9.5$ & $9.10$ & $2286$ & $9.60\pm 0.04$\\
    $9.5<\log_{10}(M_{\star}/{\rm M_{\odot}})<10.5$ & $10.06$ & $1796$ &$9.94\pm 0.02$\\
    $\log_{10}(M_{\star}/{\rm M_{\odot}})>10.5$ & $10.85$ & $494$ & $10.12\pm 0.03$\\
    \midrule
    & {\bf CHILES} & \\
    \midrule
    $8.0<\log_{10}(M_{\star}/{\rm M_{\odot}})<9.5$ & $9.10$ & $1221$ &$9.55\pm 0.07$\\
    $9.5<\log_{10}(M_{\star}/{\rm M_{\odot}})<10.5$ & $10.03$ & $767$ &$9.82\pm 0.06$\\
    $\log_{10}(M_{\star}/{\rm M_{\odot}})>10.5$ & $10.83$ & $142$ & $10.13\pm 0.08$\\
   \bottomrule
   \end{tabular}
    \vspace{1mm}
    \caption{Data points from the \HI{} spectral stacking analysis performed by \cite{Bianchetti2025}. The first column reports the $M_\star$ interval underlying each of the measured data point, the second column the corresponding median $M_\star$, the third column the number of stacked galaxies, and the fourth column the average $M_{\rm HI}$ measured from stacking.}
    \label{tab:stacking_results}
\end{table*}
    

\section{Reference cosmological hydrodynamic simulations}\label{sec:refsim}

In this section, we summarize the main features of the two cosmological hydrodynamic simulation that we use in this paper for comparison with our results, \texttt{SIMBA} \citep{Dave2019} and \texttt{IllustrisTNG} \citep{Nelson2019}. In what follows we describe only the pieces of information relevant for this work, and refer to \cite{Dave2019} and \cite{Nelson2019} for a thorough description of the simulations.

\subsection{SIMBA}

\texttt{SIMBA} is a cosmological hydrodynamic simulation run with the \texttt{GIZMO} meshless finite mass hydrodynamics. It was run in three different combinations of box sizes and resolutions: with $2\times N_{\rm dm}=1024^3$ particles (namely, $N_{\rm dm}=1024^3$ dark matter particles and $N_{\rm gas}=1024^3$ gas elements) in a $V=(100\,{\rm Mpc}\,h^{-1})^3$ comoving volume (hereafter \texttt{SIMBA-100}), $2\times N_{\rm dm}=512^3$ in a volume $V=(50\,{\rm Mpc}\,h^{-1})^3$ (hereafter \texttt{SIMBA-50}), and $2\times N_{\rm dm}=512^3$ in a volume $V=(25\,{\rm Mpc}\,h^{-1})^3$, i.e. at higher resolution with respect to \texttt{SIMBA-100} and \texttt{SIMBA-50}. The \texttt{SIMBA} fiducial model adopts and updates star formation and feedback sub-grid prescriptions used in the \texttt{MUFASA} simulation \citep[][]{Dave2016}, and introduces the treatment of black hole growth and accretion from cold and hot gas. Moreover, models for on-the-fly dust production, growth, and destruction, and \HI{} and H$_2$ abundance computation are implemented  \citep[see][ and references therein for details]{Dave2019}. \texttt{SIMBA} has been shown to reproduce several observables, among which the galaxy stellar mass functions at $z<6$ and the $M_*-{\rm SFR}$ main sequence. In addition to the full feedback model, multiple runs of \texttt{SIMBA-50} were performed, adopting variations around the full feedback model. The different feedback variations will be described directly in Sect. \ref{sec:results:sims}. The \HI{} masses in \texttt{SIMBA} were computed by assigning all the gas in the halo to its most bound galaxy within the halo \citep{Dave2020} and by assuming the subgrid model by \cite{KrumholzGnedin2011}. In this model, the $H_2$ fraction is computed based on the local metallicity and gas column density, modified to account for variations in
resolution \citep{Dave2016}. The \HI{} fraction is then computed by subtracting off the H$_2$ fraction from the total cold gas.

In this work, we make use of the publicly-available galaxy catalogs at $z\sim 0.365$ for \texttt{SIMBA-100} and for all the \texttt{SIMBA-50} boxes. Consistently with the observational sample selection by \cite{Bianchetti2025}, we select only star-forming galaxies with $\log_{10}(M_*/{\rm M}_{\odot})>8$ by separating the star-forming and quenched populations in the $M_*-{\rm SFR}$ plane. In particular, after a detailed  inspection of the $M_*-{\rm SFR}$ plane, we heuristically flag as star-forming those galaxies with $\log_{10}({\rm SFR}/({\rm M}_\odot \, {\rm yr}^{-1}))>0.9\times\log_{10}(M_*/{\rm M}_\odot)-9.2$ at $z\sim 0.365$ \citep[see,][]{Sinigaglia2022}.

\subsection{IllustrisTNG}

\texttt{IllustrisTNG} is a suite of cosmological, gravo-magnetohydrodynamical simulations run with the moving-mesh code \texttt{AREPO} \citep{Springel2010}. \texttt{IllustrisTNG} consists of an improved and updated version of the original \texttt{Illustris} simulations suite \citep{Nelson2015}, and includes the following baryon physics mechanisms, among others: primordial and metal-line radiative cooling with an ionizing background radiation field,  stochastic star formation, an effective equation of state model accounting for the two phases of the interstellar medium, evolution of stellar populations, stellar feedback, seeding and growth of supermassive black holes, AGN feedback (and in particular, energy release from high- and low-accretion rates), and the treatment of magnetic fields.

In this work, we use the following larger-volumes runs: \texttt{IllustrisTNG-100}, run with $N=2\times 1820^3$ resolution elements and in volume $V=(75~h^{-1}~{\rm Mpc})^3$, and \texttt{IllustrisTNG-300}, run with $N=2\times 2500^3$ resolution elements and in volume $V=(205~h^{-1}~{\rm Mpc})^3$.

As estimates for the \HI{} masses, we use the catalogs released by \cite{Diemer2018}. Therein, the authors performed a detailed splitting of the cold gas into its atomic and molecular phases, by adopting different subgrid models. Out of the models tested in \cite{Diemer2018}, we choose to use the values for $M_{\rm HI}$ obtained through the model by \cite{GnedinKravtsov2011}, which was calibrated onto high-resolution simulations designed to explicitly solve for radiative transfer on the fly during the simulation. Specifically, we use the catalogs at $z=0.5$, i.e. the closest redshift available to our data. Even though it does not match perfectly the redshift we are working at, it should provide a good qualitative proxy for the HIMF at intermediate redshift such as the one from this work.

We select star-forming galaxies from \texttt{IllustrisTNG} using the same criterion used for \texttt{SIMBA}, for consistency. This is a safe choice, since the two simulations have been separately shown to reproduce well the SFR-$M_\star$ plane.


\section{Methods}\label{sec:methods}

In this work, we largely follow the methodology pioneered by \cite{Bera2022}. In particular, we aim at deriving semi-empirical constraints on the \HI{} mass function and the \HI{} cosmic density parameter $\Omega_{\rm HI}$ passing from stellar mass to a \HI{} mass by means of a $M_{\rm HI}-M_\star$ scaling relation. In this way, we leverage the well-studied evolution with redshift of the stellar mass function and only assume a scaling relation measured via spectral stacking.   

In particular, we adopt the following forward-modeling:
\begin{itemize}
    \item we consider a complete stellar mass sample from the COSMOS2020 stellar mass function at $0.2<z<0.5$ tabulated in \cite{Weaver2022};
    \item we assume the $M_{\rm HI}-M_\star$ scaling relation measured via \HI{} spectral stacking from the combination of MIGHTEE and CHILES data sets and presented in \cite{Bianchetti2025}. As will be discussed in more detail later on in the paper, this step requires an adequate assumption on the scatter of the $M_{\rm HI}-M_\star$ relation. In fact, while the stacking procedure yields the average $M_{\rm HI}$ in each of the studied stellar mass bins, it does not provide any information about the scatter;
    \item once we have converted the stellar mass sample into a \HI{} mass sample, it is straightforward to build the \HI{} mass function. 
    \item we fit a Schecther function to the \HI{} mass function, for all values of $M_{\rm HI}$ above the estimated $M_{\rm HI}$ completeness limit;
    \item we integrate the \HI{} mass function and compute the total \HI{} density $\rho_{\rm HI}$ and then compute $\Omega_{\rm HI}=\rho_{\rm HI}/\rho_c$, where $\rho_c$ is the critical density of the Universe. 
\end{itemize}

In what follows, we describe in detail each of the points outlined above.

\subsection{The reference stellar mass function}\label{sec:smf}


As anticipated in Sect. \ref{sec:cosmos2020}, we adopt the stellar mass function for star-forming galaxies at $0.2<z<0.5$ from the COSMOS2020 photometric catalog \citep{Weaver2023}. In particular, we take advantage of the publicly-available posterior distributions of the parameters for the fitted double Schechter profile as follows.

For each stellar mass function obtained by substituting the parameters from the posteriors into the model, we draw a 'synthetic' galaxy sample at $\log_{10}(M_\star/{\rm M}_\odot)>8$, i.e. the same lower $M_\star$ limit used by \cite{Bianchetti2025}. To adequately account for the uncertainty coming from the limited sample size, we fix it to be the same as the ones from the samples studied in \cite{Bianchetti2025}. Henceforth, we sample from the stellar mass function $N=6,598$ galaxies for the analysis based on the full stacking results from the combination of MIGHTEE and CHILES. We adopt $N=4,576$ galaxies for the analysis based on MIGHTEE results, and $N=2,022$ for the analysis based on CHILES results when studying the consistency of the HIMF derived separately from the two surveys in Appendix \ref{app:survey_consistency}. In addition, we explore the impact of the assumed sample size in Appendix \ref{app:sample_size}.

\subsection{Converting stellar masses into HI masses}\label{sec:methods:mstar_to_mhi}

To pass from the stellar mass sample obtained as described in Section \ref{sec:smf}, we adopt a scaling relation from literature results. In particular, we adopt the most recent findings from \cite{Bianchetti2025}. As described in Sect. \ref{sec:ref_data}, therein the authors performed \HI{} stacking in $4$ stellar mass bins for the combined case, and $3$ stellar mass bins separately for MIGHTEE and CHILES. For the different cases, the authors fitted a single power law model of the type $\log_{10}(M_{\rm HI}/{\rm M}_\odot)=b \log_{10}(M_\star/{\rm M}_\odot)+c$, with $b$ and $c$ free parameters, i.e. a linear function in logarithmic space (hereafter the {\it linear} model). In this work, we generalize this framework by fitting also a second-order polynomial in logarithmic space, parametrized as $\log_{10}(M_{\rm HI}/{\rm M}_{\odot})=a \log_{10}(M_\star/{\rm M}_{\odot})^2+ b \log_{10}(M_\star/{\rm M}_{\odot})+c$ (hereafter the {\it polynomial} model). This higher-order polynomial allows to capture a potential deviation from a single power law model --- i.e. a linear model in logarithmic space --- which has been shown to be the case for results at $z\sim 0$ based on direct \HI{} detection \citep[see e.g.][]{Huang2012,Maddox2015,Parkash2018}. Such a phenomenon is typically described by means of a functional form which decreases more quickly towards low $M_{\rm HI}$ and flattens toward high $M_{\rm HI}$ values. To account for this behaviour, the following double power law model was proposed \citep[e.g.][]{Maddox2015,Parkash2018,Pan2023}:
\begin{equation}
\log_{10}(M_{\rm HI}/{\rm M}_{\odot}) = \log_{10}\left(\frac{M_0/{\rm M}_{\odot}}{\left(\frac{M_\star}{M_{\rm tr}}\right)^\alpha + \left(\frac{M_\star}{M_{\rm tr}}\right)^\beta}\right) \quad ,  
\end{equation}
with $M_0$, $M_{\rm tr}$, $\alpha$ and $\beta$ free parameters. We notice that $\alpha$ and $\beta$ denote the low- and high-mass end slopes and that this model reduces to the single power law model when $\alpha=\beta$. In this work, since we have at most $4$ data points available for the combined stacking case, we cannot fit this model --- which has $4$ free parameters --- but at most a model with $3$ free parameters, otherwise we would be overfitting our data. This motivates further the choice of a second-order polynomial, as alternative to a linear fit. Later on in the paper, we will show that the best-fitting polynomial model qualitatively reproduces the same trend as the double power law model discussed above. We notice that for the cases of the stacking performed separately on MIGHTEE and CHILES, for which we have only $3$ data points each, we are again in a situation where a full fit of a second-order polynomial with $3$ free parameter would result in overfitting. We will discuss this issue in Section \ref{sec:results:fit_mhi_mstar} and show how we reduce the number of free parameters from $3$ to $2$ to have a well-defined  fit also in the MIGHTEE and CHILES separate cases.

We estimate the best-fitting parameters for the linear and polynomial models and the related uncertainties by means of a bootstrap resampling (see \cite{Bianchetti2025}). While this has already been performed by \cite{Bianchetti2025} for the linear model, it had not been done for the polynomial one. In particular, we resample the data points underlying the fit by randomly-sampling a lognormal uncertainty based on the uncertainty on the estimation of $M_{\rm HI}$ of each data point.\footnote{We caution that the uncertainties on the $M_{\rm HI}$ estimates, obtained via jackknife resampling by \cite{Bianchetti2025}, refer to the error on the measurement and they do not represent the full underlying $M_{\rm HI}$ distribution. In particular, the jackknife uncertainty typically underestimates the true dispersion.} For each bootstrap sample, we then apply the full forward model and obtain the HIMF. 

In addition to the best-fitting $M_{\rm HI}-M_\star$, which provides only the average trend, we also need to model the scatter of the relation. Because the scatter is not measurable via traditional stacking, which yields only the average $M_{\rm HI}$ value in each stellar mass bin but not the underlying dispersion, we need to make an assumption of a credible value for the scatter. \cite{Bera2022} showed that properly modelling the scatter is of crucial importance to capture the behaviour of the HIMF. Therein, the authors performed stacking in $M_B$ intervals and derive the HIMF by means of the $M_{\rm HI}-M_B$ relation. They assumed a scatter $\sigma = 0.26$ dex -- the same as in the Local Universe --- which was proven to correctly reproduce the HIMF from ALFALFA at $z\sim 0$. 

In this work, since we are using a different scaling relation, we also need to assume a different value for the scatter. In particular, we make the same assumption as the one from \cite{Bera2022} and use the value of the scatter at $z\sim 0$ from the $M_{\rm HI}-M_\star$ scaling relation measured by the xGASS survey \citep{Catinella2018}, i.e. $\sigma=0.39$ dex. We study the specific impact of the scatter in our framework in Appendix \ref{app:scatter}, where we show that the value for the scatter largely affects the high-mass end of the HIMF. Specifically, the high-mass end of the HIMF is shown to increase with increasing value of the scatter. 
Therefore, we incorporate an additional source of uncertainty on the scatter as follows. For each step of our Monte Carlo procedure, we add to the scatter an error randomly-sampled from a Gaussian distribution with zero mean and standard deviation $\sigma_{\rm err}=0.05$. The value for this additional uncertainty contribution is motivated by covering the range of values $\sigma=(0.39\pm 0.05)$ dex, i.e. a conservatively wide range of credible values that the scatter can assume according to $z=0$ observations. In this way, we self-consistently account for the error on the scatter in each of the sampling steps and propagate the related uncertainty throughout our forward model. In Appendix \ref{app:scatter}, we show that this source of stochasticity yields no significant difference on the uncertainties on the best-fitting parameters of the HIMF with respect to the case where no scatter is assumed, nor introduces any obvious systematic shift in the median values. Therefore, in the remainder of the paper we assume as fiducial configuration the one including the error on the scatter, even though we argue that this has been shown to have little impact. In addition, in Appendix \ref{app:asymmetric} we study the impact of relaxing the assumption of a Gaussian symmetric scatter in log space, and test whether adopting a more realistic asymmetric functional form has any impact on the final results. Specifically, we model the scatter as a lognormal distribution in log space with parameters determined by fitting the true \HI{} distribution at fixed $M_\star$ as measured by the cross-matched SDSS-ALFALFA catalog \cite{Durbala2020}. We find that accounting for an asymmetric scatter has no significant impact with respect to assuming a symmetric scatter, and stick to the simpler Gaussian model for simplicity.

At the end of this procedure, we have obtained a sample of HIMFs, from which we compute the binwise median and standard deviation, representing respectively the final HIMF and its associated uncertainty. This defines our `measured' HIMF.

\subsection{Fitting the HI mass function and derivation of $\Omega_{\rm HI}$}\label{sec:methods:fit_himf}

After having obtained the average HIMF and the associated uncertainty as described in Section \ref{sec:methods:mstar_to_mhi}, we fit a single Schechter profile to it. 

In order to perform an unbiased estimate of the best-fitting parameters, we first establish the minimum $M_{\rm HI}$ that we can trust given our model, which is equivalent to the observed completeness limit. While it is not trivial to establish such a limit as classical methods  would not be applicable in the present case, we limit ourselves to a visual inspection of the resulting HIMF. We discuss this in Appendix \ref{app:completeness}. Fig. \ref{fig_himf_completeness} reveals that at $\log_{10}(M_{\rm HI}/{\rm M}_\odot)\lesssim 9$ the HIMF shows clear signs of incompleteness. Based on this visual estimate, we conservatively fit the Schechter model only at $\log_{10}(M_{\rm HI}/{\rm M}_\odot)> 9.2$ for the polynomial model and at $\log_{10}(M_{\rm HI}/{\rm M}_\odot)> 9.4$ for the linear model.

We fit a single Schechter profile to the resulting HIMF with the following parametrization\footnote{Notice that the parametrization here is slightly different from the one used in Eq. \ref{eq:double_schechter}. In Eq. \ref{eq:single_schechter} we use the same as in \cite{Ponomareva2023}, while in Eq. \ref{eq:double_schechter} we use the one from \cite{Weaver2023} for consistency with their work.}:
\begin{equation}\label{eq:single_schechter}
    \Phi (M_{\rm HI}) = \log(10)~\tilde{\Phi}\left(\frac{M_{\rm HI}}{\tilde{M}}\right)^{\alpha+1}e^{\left(-M_{\rm HI}/\tilde{M}\right)}
\end{equation}
where the free parameters $\alpha$, $\tilde{\Phi}$, and $\tilde{M}$ represent the faint-end slope, the normalization, and the `knee' mass of the mass function. For the fitting procedure, we adopt a Markov Chain Monte Carlo approach and use the affine-invariant \texttt{emcee} ensemble sampler \citep{ForemanMackey2013}. For each of the tested models, we run $50,000$ steps and discard the first $1,000$ to conservatively avoid the burn-in phase. We then estimate the best-fitting value as the median ($50th$ percentile) and the associated uncertainty as difference ($84th-50th$) percentile (upper) and ($50th-16th$) percentile (lower).  

Once we have obtained the best-fitting values for the parameters of the Schechter model, we integrate the HIMF to derive the \HI{} cosmic density $\rho_{\rm HI}$. In particular, the integral of the Schechter function can be expressed in terms of the incomplete Euler gamma function $\Gamma$\footnote{We use $\Gamma$ instead of $\gamma$ to denote the incomplete Euler Gamma function for consistency with other references.} as:
\begin{equation}
    \rho_{\rm HI}=\Gamma(\alpha+2)\tilde{\Phi} \tilde{M} \quad .
\end{equation}

To adequately obtain the uncertainty on $\rho_{\rm HI}$, we compute the integral for all the samples from the posterior distributions, build the resulting distribution of values of $\rho_{\rm HI}$, and compute the best-fitting and related uncertainty using again the $16th$, $50th$, and $84th$ percentiles as described above.

Eventually, the \HI{} cosmic density parameter $\Omega_{\rm HI}$ is calculated as:
\begin{equation}
    \Omega_{\rm HI} = \frac{8\pi G}{3H_0^2}\rho_{\rm HI} \quad ,
\end{equation}
where $\rho_c=3H_0^2/(8\pi G)$ is the critical density of the Universe. We notice that $\Omega_{\rm HI}$ depends on the choice of $H_0$. Nonetheless, where relevant for the comparison we have rescaled the other results from the literature to the same $H_0$ value.

\section{Results}\label{sec:results}

In this section, we present the results of our analysis and their discussion.

\subsection{Fitting the $M_{\rm HI}-M_\star$ relation}\label{sec:results:fit_mhi_mstar}

\begin{figure}
    \centering
    \includegraphics[width=\columnwidth]{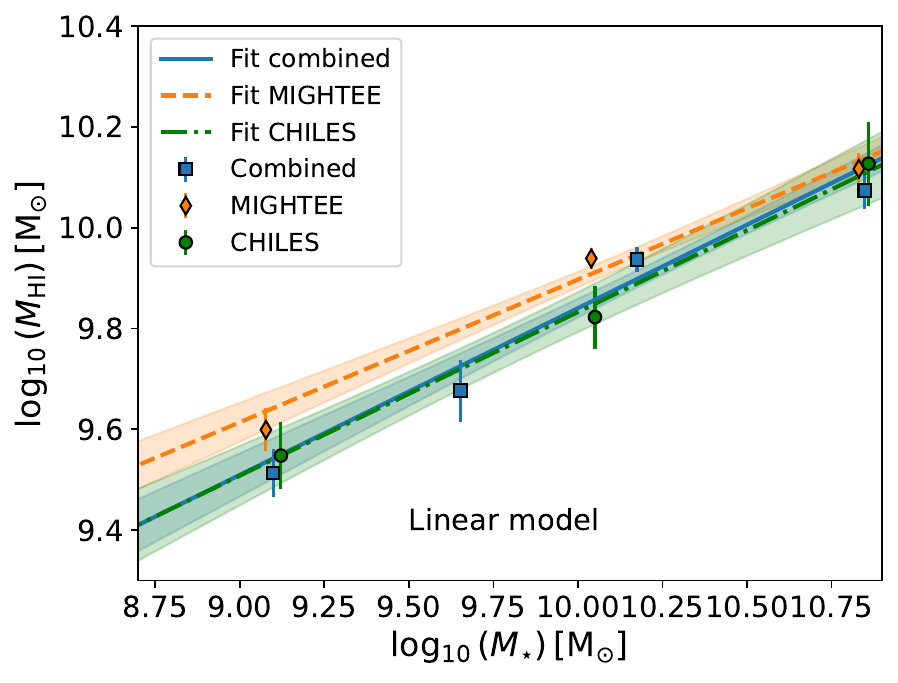}
    \includegraphics[width=\columnwidth]{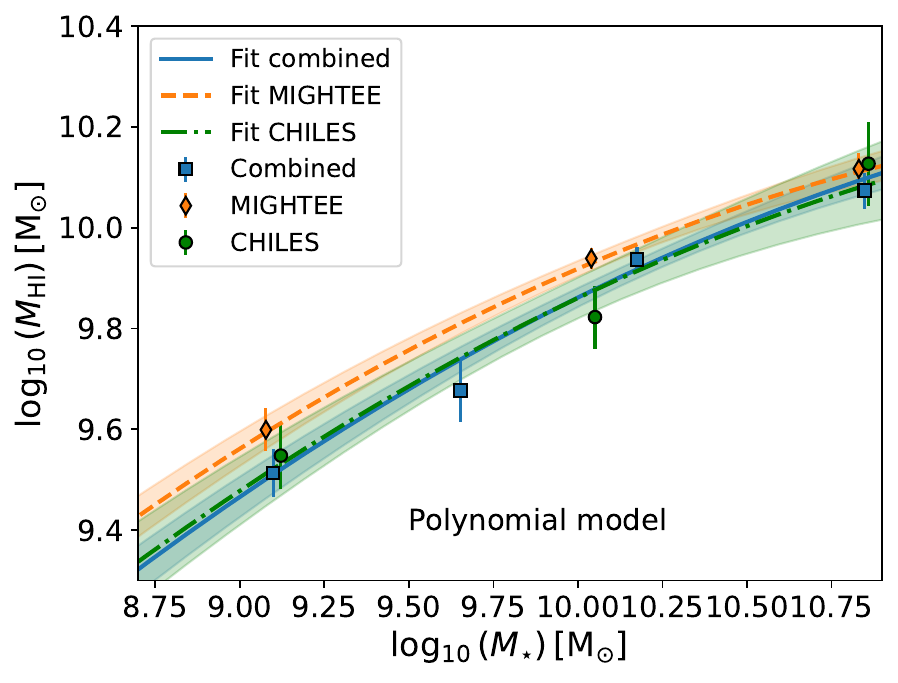}
    \caption{$M_{\rm HI}-M_\star$ best-fitting scaling relations to the stacking results from \cite{Bianchetti2025}. We show the linear fit and the polynomial fit to the data in the top and bottom panel, respectively. The data points are shown as blue squares (combined stacking), green diamonds (MIGHTEE stacking), and green circles (CHILES stacking). The best fit models are displayed as blue solid (combined stacking), orange dashed (MIGHTEE stacking), and green dotted-dashed lines (CHILES stacking), respectively.}
    \label{fig:mhims_relation}
\end{figure}

Fig. \ref{fig:mhims_relation} shows the resulting fit of the data points from \cite{Bianchetti2025} for the linear model (top panel) and polynomial model (bottom panel). We show the data from the combined stacking as blue squares and display the best fit model to the data as a blue solid line. In addition, to check the consistency of the fit of the scaling relation, we also show the data from MIGHTEE and CHILES as orange diamonds and green circles, respectively, as well as the best-fitting models as an orange dashed line (MIGHTEE), and a green dotted-dashed (CHILES).

When fitting the polynomial model to the combined case, we find that the free parameter $c$ defined in Section \ref{sec:methods:mstar_to_mhi} is not well constrained and consistent with zero. If we fix $c=0$, we find that the best-fitting values for $a$ and $b$ remain unchanged, but the resulting uncertainties on both parameters are much smaller. Therefore, we decide to fix $c=0$ for all the polynomial fits. This has two main advantages: (i) it lifts the parameter degeneracy discussed above, and (ii) it allows to have a well-defined fit also in the MIGHTEE and CHILES cases. 

We report the best-fitting parameter values for both the linear and the polynomial model in Table \ref{tab:bestfit_relation}. In particular, we leave blank the column of the values for $a$ for the linear fit because the model does not have the second-order term, and the column of the values for $c$ in the polynomial fit because we have set $c=0$, as discussed above.

From Fig. \ref{fig:mhims_relation}, we see that both the linear and the polynomial models provide reasonable fits to the data. We compute the 'reduced chi square', $\chi^2/({\rm degrees~of~freedom})$, and report the values in the last column of Table \ref{tab:bestfit_relation}. Given the low number of degrees of freedom, it is hard to meaningfully establish which among the two models is preferred. Therefore, we decide to adopt as fiducial model the polynomial one which, as anticipated in Section \ref{sec:methods:mstar_to_mhi}, having $a<0$ has the same qualitative properties as a double power law model, i.e. it decreases more rapidly than a linear model towards low \HI{} masses and flattens towards high \HI{} masses.

\begin{table}[]
    \centering
    \scriptsize
    \begin{tabular}{lcccc}
    \toprule
       Survey & $a$ & $b$ & $c$ & 
       $\chi^2/{\rm d.o.f.}$ \\
       \midrule
       & &  Lin. fit & & \\
       \midrule
       MIGHTEE &  - & $0.286\pm 0.032$ & $7.029\pm 0.290$ & $2.764$ \\
       \addlinespace[1.5mm]
       CHILES & - & $0.328\pm 0.065$ & $6.547\pm 0.607$ & $0.174$\\
       \addlinespace[1.5mm]
       Combined & - & $0.326\pm 0.030$ & $6.579\pm 0.339$ & $2.670$ \\
       \midrule
        & & Poly. fit & & \\
       \midrule
       MIGHTEE &  $-0.070\pm 0.002$ & $1.693\pm 0.027$ & - & $0.005$ \\
       \addlinespace[1.5mm]
       CHILES & $-0.067\pm 0.006$ & $1.651\pm 0.061$ & - & $1.043$\\
       \addlinespace[1.5mm]
       Combined & $-0.066\pm 0.003$ & $1.648\pm 0.064$ & - & $1.002$ \\
        \bottomrule
    \end{tabular}
    \vspace{1mm}
    \caption{Best-fitting values for the parameters of the model described in Sect. \ref{sec:methods:mstar_to_mhi} for the $M_{\rm HI}-M_\star$ scaling relation when fitting the data points by \cite{Bianchetti2025}. Top: best-fitting parameters of the linear fit ($a=0$ by construction). Bottom: best-fitting parameters of the polynomial fit ($c=0$ set as choice).}
    \label{tab:bestfit_relation}
\end{table}

\begin{table*}[]
    \centering
    \footnotesize
    \begin{tabular}{lccccccc}
    \toprule
       Survey   & z & $\log_{10}(\tilde{M}/{\rm M}_{\odot})$ & $\alpha$ & $\log_{10}(\tilde{\Phi}/{\rm Mpc}^{-3})$ & 
       $\Omega_{\rm HI}\times 10^{-4}$ & Reference \\
       \midrule
       ALFALFA.100 & 0 & $9.93\pm0.01\pm 0.005$ & $-1.25\pm 0.02\pm 0.1$ & $-2.28\pm 0.02\pm 0.07$ & $3.90\pm 0.10$ & \cite{Jones2018}\\
       \addlinespace[1.5mm]
       MIGHTEE $1/V_{\rm max}^{5S_{\rm lim}}$ & 0.04 &  $10.04^{+0.24}_{-0.24}$ & $-1.29^{+0.37}_{-0.26}$ & $-2.29^{+0.32}_{-0.36}$ & $5.26^{+0.91}_{-0.95}$ & \cite{Ponomareva2023}\\
       \addlinespace[1.5mm]
       MIGHTEE $1/V_{\rm max}^{8S_{\rm lim}}$ & 0.04 &  $10.07^{+0.22}_{-0.24}$ & $-1.40^{+0.42}_{-0.24}$ & $-2.31^{+0.35}_{-0.38}$ & - & \cite{Ponomareva2023}\\
       \addlinespace[1.5mm]
       MIGHTEE MML & 0.04 &  $10.19^{+0.10}_{-0.13}$ & $-1.44^{+0.13}_{-0.10}$ & $-2.47^{+0.19}_{-0.14}$ & $6.08^{+0.30}_{-0.30}$ & \cite{Ponomareva2023}\\
       \addlinespace[1.5mm]
       AUDS & 0.16 &  $10.14\pm 0.09$ & $-1.37\pm 0.05$ & $-2.55\pm 0.01$ & $3.55\pm 0.29$ & \cite{Hoppman2015}\\
       \addlinespace[1.5mm]
        BUDHIES & 0.2 &  $9.77^{+0.16}_{-0.04}$ & $-1.49\pm 0.48$ & $-2.25\pm 0.03$ & $4.10\pm 4.43$ & \cite{Gogate2022}\\
       \addlinespace[1.5mm]
       HI int. mapping & 0.32 & $9.46^{+0.38}_{-0.31}$ & $-1.29^{+0.36}_{-0.39}$ & $-1.80^{+0.65}_{-0.72}$ & $4.13^{+5.55}_{-1.58}$ & \cite{Paul2023}\\
       \addlinespace[1.5mm]
       \midrule
       \midrule
       Combined (poly.) & 0.37 &  $10.16^{+0.08}_{-0.08}$ & $-1.48^{+0.10}_{-0.09}$ & $-2.41^{+0.12}_{-0.13}$ & $7.02^{+0.59}_{-0.52}$ & This work\\
       \addlinespace[1.5mm]
       Combined (linear) & 0.37 & $10.13^{+0.09}_{-0.09}$ & $-1.56^{+0.14}_{-0.13}$ & $-2.31^{+0.15}_{-0.17}$ & $9.98^{+1.78}_{-1.12}$ & This work\\
        \bottomrule
    \end{tabular}
    \vspace{1mm}
    \caption{Numerical results for the best-fitting parameters to the HIMF using a Schechter model as in Eq. \ref{eq:single_schechter}, for the results from this work and from different literature results. The first column reports the surveys the results refer to; the second column reports the median redshift at which the HIMF has been measured; the third, fourth and fifth columns report the best-fitting values for the $\log_{10}(\tilde{M}/{\rm M}_{\odot})$, $\alpha$, and $\log_{10}(\tilde{\Phi}/{\rm Mpc}^{-3})$; the sixth column reports the inferred value for $\Omega_{\rm HI}$; the seventh column displays the reference presenting the results.}
    \label{tab:results_himf}
\end{table*}

\subsection{The HI mass function and $\Omega_{\rm HI}$ at $z\sim 0.37$}\label{sec:results:himf}

We fit a Schechter model to the resulting HIMF for the combined stacking and for both models for the $M_{\rm HI}-M_\star$ scaling relation (linear and plynomial) as described in Section \ref{sec:methods:fit_himf}. We report in Table \ref{tab:results_himf} the results for the best-fitting values of the Schechter parameters $\{ \tilde{M}, \alpha, \tilde{\Phi}\}$, as well as the resulting value for $\Omega_{\rm HI}$ computed as described in Sect. \ref{sec:methods:fit_himf}, together with a compilation of results from the literature. We report the results from the fitting of the HIMF resulting from MIGHTEE and CHILES stacking separately in Appendix \ref{app:survey_consistency}.

Fig. \ref{fig:himf_global} shows the resulting HIMF from this work as yellow stars (linear model) and blue squares (polynomial model). We display as blue and yellow solid lines the best-fitting Schechter profiles as found from the median of the posterior distributions of the parameters. We also display the results for the HIMF from the literature reported in Table \ref{tab:results_himf}. In particular, we show the results from ALFALFA at $z\sim 0$ \citep[][light green dashed]{Jones2018}, from MIGHTEE at $0< z < 0.084$ \citep[][red dotted and gray dotted-dashed]{Ponomareva2023}, from the AUDS \citep[][brown dotted-dashed]{Hoppman2015} and BUDHIES \citep[][cyan dashed]{Gogate2022} surveys at $z\sim 0.16$ and $z \sim 0.2$, respectively (all based on direct detections), from a HI intensity mapping experiment at $z\sim 0.32$ \citep[][purple dotted-dashed]{Paul2023}, and green circles and orange the semi-empirical HIMF results obtained by \cite{Bera2022} at $z\sim 0.35$ and \cite{Chowdhury2024} at $z\sim 1$ following similar approaches to the one adopted herein. We find difference between the HIMF reported in this work and the one at $z\sim 0$ from ALFALFA. In particular, the HIMF derived here with both models have a consistently higher normalization than the ALFALFA HIMF, on average by a factor $\sim 2$. Compared to the MIGHTEE results at lower redshift and to the AUDS survey at $z\sim 0.16$, the HIMF from this work has a larger normalization at low $M_{\rm HI}$ values, whereas they tend to converge towards high $M_{\rm HI}$. The results from BUDHIES, from \cite{Bera2022}, and from \cite{Paul2023} have comparable normalization to the one of our results at low $M_{\rm HI}$, but they present a rapid fall-off towards high $M_{\rm HI}$, dropping below the ALFALFA HIMF at $\log_{10}(M_\star/{\rm M_{\odot}})\lesssim 9.6$. Essentially, there is evidence for a variation between our results at $z\sim 0.37$ and the results by \cite{Chowdhury2024} at $z\sim 1$, the latter having a consistently larger normalization at the probed \HI{} masses. These differences could be due to potential evolution of the HIMF over cosmic times, or to possible selection biases or other systematics. In particular, the approach used in this work relies on a galaxy sample which is extracted from the COSMOS2020 catalog, and hence, selected in optical $M_\star$. This is not the case for e.g. blind \HI{} surveys to which we are comparing our results. In general, the mentioned \HI{} surveys feature a severe inhomogeneity in the covered area, depth and sensitivity, instrument and observational configuration (e.g., single dish vs interferometric), among others. In addition, potential systematics related to the interferometric data from MIGHTEE and CHILES may also have an impact in the final results. Finally, as mentioned earlier in this section, the results on the HIMF to which we are comparing our findings are obtained following different methods. Therefore, given these aspects altogether, it is hard to conclusively tell if the detected differences between our results and the other tabulated results are fully due to cosmic evolution or are contaminated by the effects discussed above.

\begin{figure*}
    \centering
    \includegraphics[width=\textwidth]{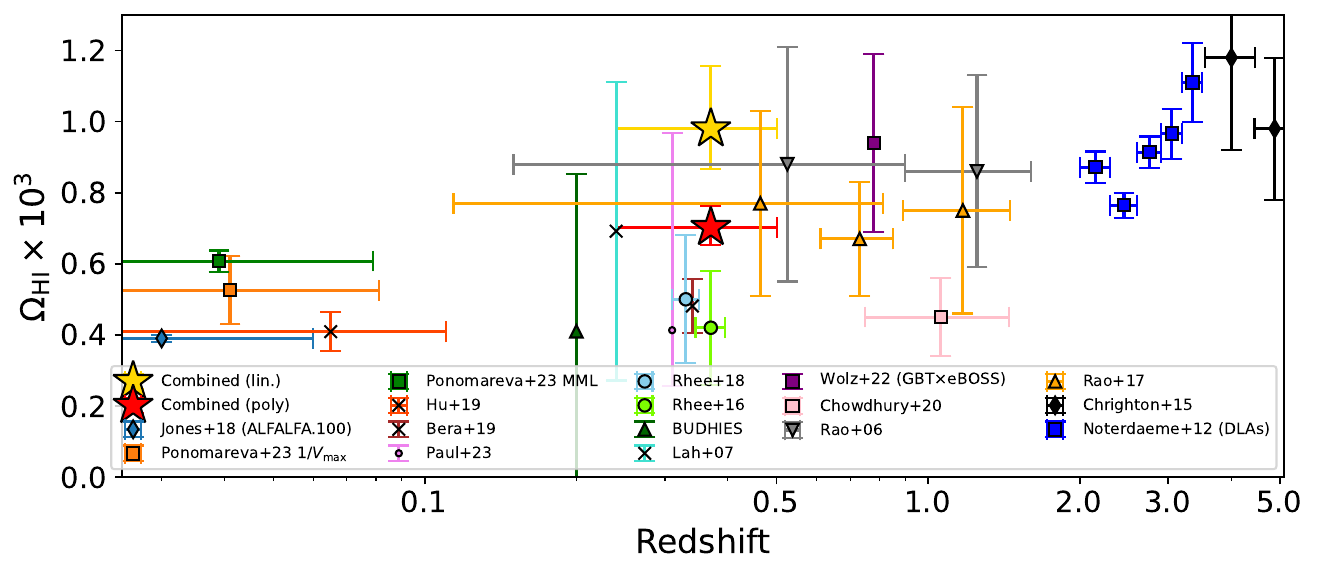}
    \caption{Compilation of $\Omega_{\rm HI}$ results as function of redshift from this work and from the literature. We display the results as follows: from this work, as stars, in particular the result from the combined stacking with a linear model in yellow and with a polynomial model in red; from ALFALFA at $z\sim 0$ as a blue diamond \citep{Jones2018}; from MIGHTEE a t $0<z<0.084$ as orange ($1/V_{\rm max}$ method) and green (MML method) \citep{Ponomareva2023}; from a \HI{} stacking based on WSRT data at $z\sim 0.066$ as a red cross \citep{Hu2019}; from a \HI{} stacking on uGMRT data at $z\sim 0.35$ as a brown cross \citep{Bera2019}; from a \HI{} intensity mapping experiment with MeerKAT as a pink circle \citep{Paul2023}; from two \HI{} stacking experiments based on uGMRT data at $\sim 0.32$ \citep{Rhee2016} and $z\sim 0.37$ \citep{Rhee2016} as blue and green circles, respectively; from BUDHIES at $\sim 0.2$ as a green upward triangle \citep{Gogate2022}; from \HI{} stacking at $z\sim 0.24$ based on GMRT data as a cyan cross \citep{Lah2007}; from the cross-correlation between \HI{} intensity mapping from the GBT and optical galaxies from the eBOSS survey as a purple square \citep{Wolz2022}; from Damped Ly$\alpha$ Systems (DLAs) at $z<1.65$ from \cite{Rao2006} and \cite{Rao2017} as downward gray triangles and orange upward triangles, respectively; from DLAs at $z>2$ \citep{Noterdaeme2012} and at $z>4.4$ \citep{Chrighton2015} as dark blue squares and black diamonds, respectively.}
    \label{fig:omegahi}
\end{figure*}

Fig. \ref{fig:omegahi} shows the results for $\Omega_{\rm HI}$ as a function of redshift derived in this work. We show the outcome from the combined stacking adopting a linear and polynomial model as a yellow and a red star respectively. In addition, we plot in the same figure a compilation of results from the literature. We notice that:
\begin{itemize}
    \item our results are in broad agreement with other literature results at the same redshift, although the scatter in these results is large and encompasses values spanning an interval $0.35\times10^{-3}\lesssim \Omega_{\rm HI} \lesssim 0.95\times 10^{-3}$;
    \vspace{0.5mm}
    \item the linear model predicts consistently larger HIMF than the polynomial model. This translates into systematically larger values for $\Omega_{\rm HI}$, as can be seen in Table \ref{tab:results_himf};
    \item the combined result adopting a polynomial model at $z\sim 0.37$ $\Omega_{\rm HI}=7.02^{+0.59}_{-0.52}\times 10^{-3}$ is compatible within $1\sigma$ with the results by MIGHTEE at $0\lesssim z \lesssim 0.084$, but deviates at $\sim 2.9\sigma$ from the value $\Omega_{\rm HI}=(3.90\pm 0.10)\times 10^{-3}$ reported by ALFALFA at $z\sim 0$;
    \item the resulting values for $\Omega_{\rm HI}$ obtained by assuming the linear and the polynomial model for the scaling relation differ at $\sim 2.4\sigma$ for the combined stacking;
    \item our results for $\Omega_{\rm HI}$ follow the overall trend for $\Omega_{\rm HI}$ increasing with increasing redshift. 
\end{itemize}

\begin{figure}
    \centering
    \includegraphics[width=\columnwidth]{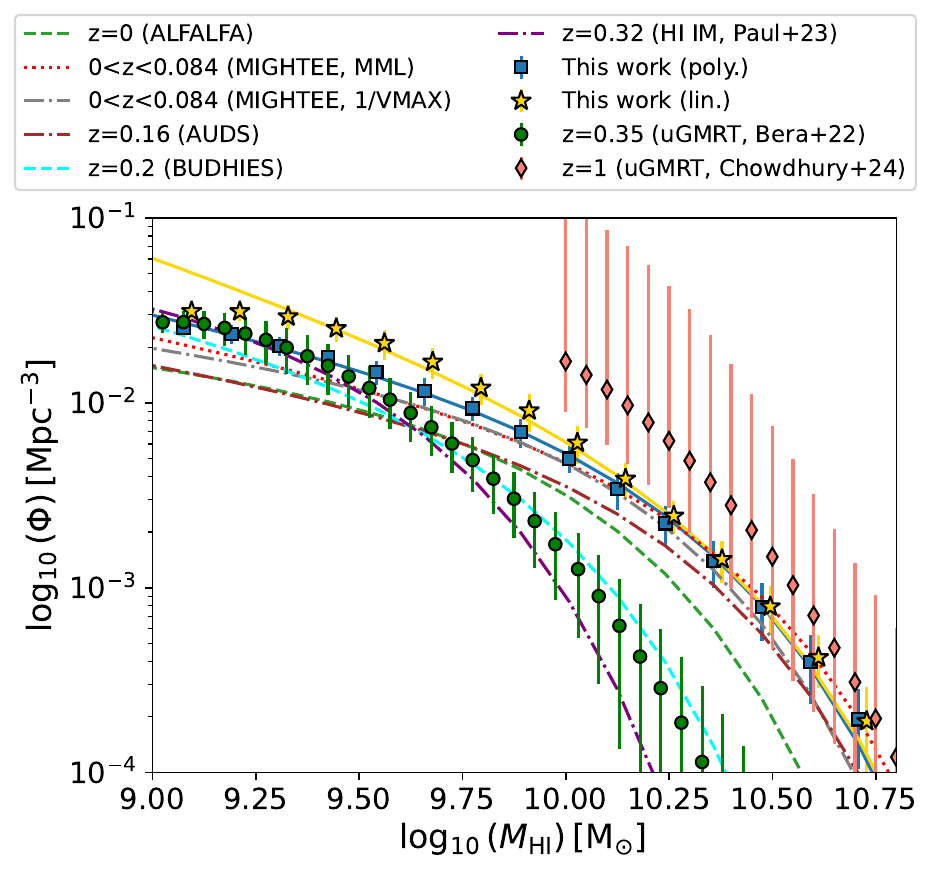}
    \caption{Compilation of HIMF results from this work and from the literature. We show as yellow stars and blue squares the resulting HIMF from the linear and polynomial models, respectively, as well as their best-fitting models as solid lines of the same color. We display the HIMF from ALFALFA at $z\sim 0$ \citep{Jones2018} as a green dashed line, from MIGHTEE at $0<z<0.084$ as red dotted ($1/V_{\rm max}$) and gray dotted-dashed (MML) lines \citep{Ponomareva2023}, from AUDS at $z\sim 0.16$ as a brown dotted-dashed line \citep{Hoppman2015}, from BUDHIES at $z\sim 0.2$ as a cyan dashed line \citep{Gogate2022}, from the intensity mapping experiment by at $z\sim 0.32$ as a purple dotted-dashed line \citep{Paul2023}, and as green circles and orange diamonds the results from a similar approach as the one followed herein applied to stacking results from uGMRT data at $z\sim0.35$ \citep{Bera2022} and $z\sim 1$ \citep{Chowdhury2024}.} 
    \label{fig:himf_global}
\end{figure}

\subsection{Discussion}\label{sec:discussion}

As already anticipated, the main findings of this work point towards a difference in the HIMF and in the $\Omega_{\rm HI}$ parameters with respect to the fiducial ALFALFA HIMF at $z\sim 0$. This fact is consistently inherited from the scaling relation from \cite{Bianchetti2025}, which also features a clear deviation with respect to literature results at $z\sim 0$. Assuming that the detected variations are due to actual evolution, and including the HIMF from \cite{Chowdhury2024} in the picture, the \HI{} content of galaxies would evolve gradually and smoothly from $z\sim 0$ to $z\sim 1$. This scenario is in tension with the results from \cite{Bera2022} and from \cite{Paul2023}. In fact, both references promote a HIMF with similar normalization at $\log_{10}(M_{\rm HI}/{\rm M}_\odot)\sim 9-9.5$ as from this work, but with a quicker fall-off towards high $M_{\rm HI}$. This typically results in a lower value for $\Omega_{\rm HI}$. In the case of \cite{Bera2022}, the results is obtained with a similar approach as ours, but the authors therein used a different scaling relation to pass from $M_B$ to $M_{\rm HI}$. In addition, the sample size is a factor $\sim 10$ smaller than the one used in \cite{Bianchetti2025}, so cosmic variance could be affecting at least partially the results when measuring \HI{} masses from stacking. Eventually, the selection of star-forming galaxies performed by \cite{Bera2022} and \cite{Bianchetti2025} are based on different criteria based on colors.  These aspects altogether may explain the observed differences, although we regard this comparison as not conclusive as in general more investigation with larger sample sizes is required. With respect to \cite{Paul2023}, the HIMF obtained in that work are based on \HI{} intensity mapping, i.e. a radically different approach from the one adopted herein. Therefore, it is difficult to establish a fair comparison in this case. Nonetheless, it is worth mentioning that the result obtained via \HI{} intensity mapping is in good agreement with the HIMF obtained by \cite{Bera2022}.


We notice that our results have a non-negligible dependence on the model adopted to describe the $M_{\rm HI}-M_\star$ relation. In fact, the linear model tends to produce HIMF with higher normalization and larger values for $\Omega_{\rm HI}$ than the polynomial model. This is not surprising since, as discussed, the polynomial model steepens towards low $M_\star$ where a greater number of galaxies is found. In addition, a linear model extrapolates the $M_{\rm HI}-M_\star$ relation to arbitrarily high value of $M_{\rm HI}$. However, results at $z\sim 0$ show that massive galaxies feature a flattening in $M_{\rm HI}$ at the high-mass end of the $M_{\rm HI}-M_\star$ relation, due to a plethora of phenomena shaping galaxy evolution, mainly gas depletion via star formation. In this work, we find evidence for the polynomial model to better fit the data points from stacking. In particular, the best-fitting second-order polynomial model for the $M_{\rm HI}$ relation is found to have second-order coefficient $a<0$ at a significance $>10\sigma$ in all the studied cases. While we regard this finding as non-conclusive given the limited amount of data points available for the fit, the trend suggested by the best-fitting polynomial fit qualitatively reproduces the one described by a double power law at $z\sim 0$ \citep{Maddox2015,Parkash2018,Pan2023}. In particular, the polynomial model with $a<0$ accounts for the high-mass flattening of the $M_{\rm HI}-M_\star$ scaling relation. For all these reasons we regard the polynomial model as the fiducial model for this work. 



\subsection{Comparison with simulations}\label{sec:results:sims}

We also compare our findings to the predictions from publicly-available cosmological hydrodynamic simulations. In particular, we consider the \texttt{SIMBA} and \texttt{IllustrisTNG} simulations (previously described in Section \ref{sec:refsim}), and construct the HIMF for the different available models. Specifically, for the \texttt{SIMBA-50} simulation, we consider both the fiducial model and a number of alternatives varying the feedback model. 

\begin{figure}
    \centering
    \includegraphics[width=\columnwidth]{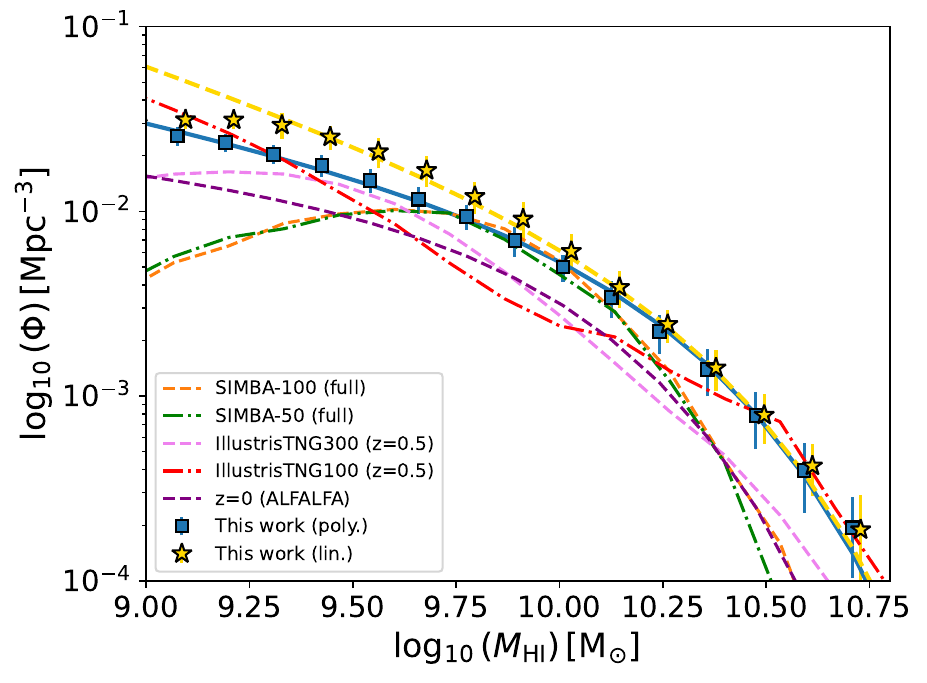}
    \includegraphics[width=\columnwidth]{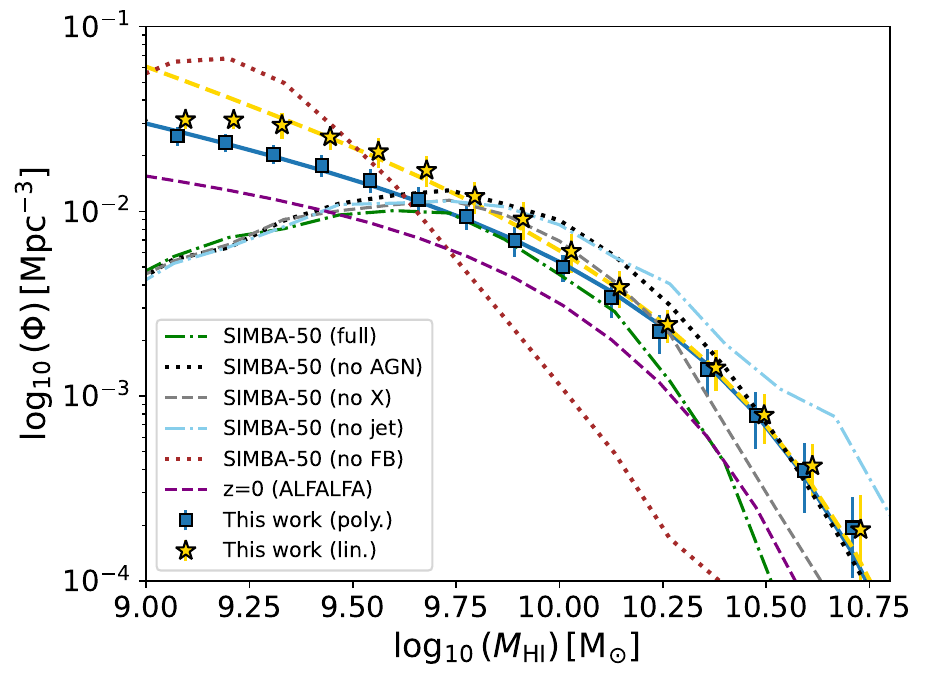}
    \caption{Comparison of the resulting HIMF from this work with the prediction from cosmological hydrodynamic simulations. Top: Comparison with the \texttt{SIMBA} full feedback model at $z\sim 0.365$ and \texttt{IllustrisTNG} at $z=0.5$. We display as yellow stars and blue squares the resulting HIMF from this work, from assuming a linear and polynomial model for the $M_{\rm HI}-M_\star$ relation. respectively. We also show as yellows dashed and blue solid lines the best-fitting Schechter functions, respectively. We display as orange dashed and green dotted-dashed lines the predictions from \texttt{SIMBA-100} and \texttt{SIMBA-50}, respectively, and with a pink dashed and red dotted-dashed line the predictions from \texttt{IllustrisTNG-300} and \texttt{IllustrisTNG-100}, respectively. We show as a puple dashed line the HIMF from ALFALFA at $z\sim 0$, for comparison. Bottom: Comparison with the different variations of the \texttt{SIMBA-50} simulations. We display as a green dashed line the prediction from the full feedback run, as a black dotted line the one from the {\it noAGN} run, as a gray dashed line the one from the {\it noX} run, as a cyan dotted-dashed line the one from the {\it noJet} run, and as a brown dotted line the one from the {\it noFB} run.} 
    \label{fig:simulations}
\end{figure}

Figure \ref{fig:simulations} shows the detailed comparison of our results with the aforementioned simulations. We show in the top panel the global comparison with \texttt{SIMBA} and \texttt{IllustrisTNG}, while in the bottom panel we focus specifically on the different feedback model implemented in \texttt{SIMBA}. We also plot the HIMF from ALFALFA at $z\sim0$ for comparison.

From the global comparison in the top panel, it turns out that there is no qualitative difference between the HIMF from \texttt{SIMBA-50} and \texttt{SIMBA-100}. The incompleteness of the HIMF at $\log_{10}(M_\star/{\rm M_\odot})\lesssim 9.7$ --- stemming from the simulation resolution --- is the main driver of the trend at low $M_{\rm HI}$, while no significant effects from cosmic variance are observed at high masses.

While at $z= 0$ both \texttt{SIMBA} and \texttt{IllustrisTNG} fit well the ALFALFA HIMF \citep{Dave2020}, none of them reproduce our observations at $z\sim 0.37$ very well at all the probed \HI{} masses. In particular, \texttt{SIMBA} fits our HIMF well at the smallest \HI{} masses at which the HIMF is complete, i.e. $9.75\lesssim\log_{10}(M_{\rm HI}/{\rm M_\odot})\lesssim 10$, but suffer a quick drop at higher \HI{} masses, converging towards the HIMF at $z\sim 0$ reported by ALFALFA. The high-mass end of the stellar and HIMF are known to be dominated by the effect of AGN feedback \citep[e.g.,][]{Beckman2017}. Therefore, it is crucial to consider the effect of feedback, which will be studied later on in this section.

\texttt{IllustrisTNG-100} and \texttt{IllustrisTNG-300} have different resolutions, which is reflected into the different completeness limit observable in the predicted HIMF. While \texttt{IllustrisTNG-100} is found to be complete for all the probed \HI{} masses (i.e. down to $\log_{10}(M_{\rm HI}/{\rm M_\odot})\sim 9$), the lower-resolution \texttt{IllustrisTNG-300} is found to become incomplete already at $\log_{10}(M_{\rm HI}/{\rm M_\odot})\lesssim 9.5$. Because the values for $M_{\rm HI}$ were computed by \cite{Diemer2018} using subgrid recipes to split the atomic and molecular phases of neutral hydrogen, it is important to take into account that the different resolution is likely to have an important impact here on the prediction of the values for \HI{}. 

Both for \texttt{IllustrisTNG-100} and for \texttt{IllustrisTNG-300} we observe a good convergence to our HIMF results at low $M_{\rm HI}$ (the lowest masses at which each HIMF is complete). At high $M_{\rm HI}$, only \texttt{IllustrisTNG-100} fits well the HIMF from this work, most likely due to resolution effects. At intermediate \HI{} masses, none of them reproduces our HIMF, but rather they predict HIMF which are more in agreement with the result at $z\sim 0$ from ALFALFA.  

In the bottom panel, we analyze the effect of the different implementations of feedback based on the \texttt{SIMBA-50} run. In fact, as already mentioned previously, from the results shown in the top panel we have observed that there are no obvious volume effects in the \texttt{SIMBA} runs, and hence, we can safely perform this study based on \texttt{SIMBA-50}, for which different versions with distinct feedback models have been produced. Specifically, we study the following different cases:
\begin{itemize}
\item \textit{full}: includes the full \texttt{SIMBA} feedback model;
\item \textit{noAGN}: the full AGN feedback model is not included;
\item {\it noX}: the X-ray AGN feedback model is not included;
\item {\it noJet}: the X-ray AGN and jet feedback models are not included;
\item {\it noFB}: no star formation nor AGN feedback models are included.
\end{itemize}

We start by observing that the {\it noFB} case clearly represents an unrealistic and oversimplified case. In fact, it produces a HIMF with an excess by a factor $\sim 5.7$ of low-$M_{\rm HI}$ galaxies at $\log_{10}(M_{\rm HI}/{\rm M_\odot})\sim 9.3$ with respect to the ALFALFA HIMF, and by a factor $\sim 2.8$ with respect to our HIMF, and a lack of high-$M_{\rm HI}$ galaxies at $\log_{10}(M_{\rm HI}/{\rm M_\odot})\sim 10.35$ by a factor $\sim 3.6$ with respect to ALFALFA and by a factor $\sim 8.3$ with respect to our results. 

All the other feedback models converge well within each other and are in good agreement with our results at $\log_{10}(M_{\rm HI}/{\rm M_\odot})\sim 9.7$, but tend to diverge towards high $M_{\rm HI}$ values. 
The {\it noX} model underpredicts our results, while the {\it noAGN} is in good agreement with them, and the {\it noJet} mode overpredicts the HIMF, respectively. This indicates that, as anticipated, the specific AGN feedback model is a major driver of the massive end of the HIMF. In particular, it is interesting to notice that the {\it noAGN} model reproduces very well the HIMF from this work. The finding suggests a scenario in which neglecting the feedback from supermassive black holes allows to better model the HIMF at the probed \HI{} masses. This is consistent with the results from \cite{Gensior2023,Gensior2024}, where the authors considered tens of Milky-Way-like galaxies from the \texttt{EMP-Pathfinder} \citep{ReinaCampos2022} and \texttt{FIREbox} \citep{Feldmann2023} cosmological simulations and showed that they reproduce well the observed \HI{} properties of nearby galaxies without the need of AGN feedback. Therefore, despite the large uncertainty associated to feedback models in current cosmological hydrodynamic simulations, we argue that a detailed study of these feedback mechanisms is crucial to properly understand which physical mechanism shapes the HIMF at the different \HI{} masses.

\section{Summary and conclusions}\label{sec:conclusions}

In this work we report a novel indirect semi-empirical estimation of the HIMF and of the \HI{} cosmic density parameter $\Omega_{\rm HI}$ for star-forming galaxies at $z\sim 0.37$. In particular, we combine the most recent measurements of the stellar mass function of star-forming galaxies in the COSMOS field at $0.2<z<0.5$ \citep{Weaver2023} and of the $M_{\rm HI}-M_\star$ scaling relation of star-forming galaxies at $0.22<z<0.49$ (median redshift $z\sim 0.37$) by \cite{Bianchetti2025} from the combined \HI{} stacking analysis of two state-of-the-art deep \HI{} interferometric surveys MIGHTEE \citep{Jarvis2016} and CHILES \citep{Fernandez2013}, both covering the COSMOS field. In particular, we adopt a Monte Carlo approach following the one pioneered by \cite{Bera2022} consisting of drawing samples of the stellar mass function, generate a synthetic stellar mass sample from each stellar mass function, and convert $M_\star$ to $M_{\rm HI}$ by assuming a $M_{\rm HI}-M_\star$ scaling relation with an adequate scatter. We compute the median and the standard deviation of the sample of HIMF, which defines our `measured' HIMF. In addition, for each \HI{} mass sample, we build the corresponding HIMF and fit a Schechter profile via a Markov Chain Monte Carlo procedure using the \texttt{emcee} affine-invariant sampler \citep{ForemanMackey2013}. This allows us to determine the full posterior distributions of the best-fitting  Schechter parameters and hence, to determine not only the median values but also their associated uncertainties after having adequately propagated all the uncertainties in our forward modelling.

The main findings of this paper can be summarized as follows:
\begin{itemize}
    \item we model the $M_{\rm HI}-M_\star$ scaling relation by fitting the data points from \cite{Bianchetti2025} using both a linear and a second-order polynomial model. In particular, the second-order coefficient $a$ is found to be negative, $a<0$. The resulting model resembles the well-known `curved trend' observed in the $M_{\rm HI}-M_\star$ relation at $z=0$;
    \item we find a global difference in the normalization of the HIMF from the $z\sim 0$ to $z\sim 1$. In particular, the resulting HIMF from this work has larger normalization than the one reported by ALFALFA at $z\sim 0$, and smaller normalization than the one reported by \cite{Chowdhury2024} at $z\sim 1$. We tentatively interpret this fact as as a signature of evolution of the HIMF, although it is hard to draw conclusions in this regard given the potential selection biases and systematics that could be affecting the comparison between our findings and the other literature results;
    \item 
    the linear model typically produces a HIMF with larger normalization than the polynomial model;
    \item the combined result adopting a polynomial model at $z\sim 0.37$ $\Omega_{\rm HI}=7.02^{+0.59}_{-0.52}\times 10^{-3}$ is compatible within $1\sigma$ with the results by MIGHTEE at $0\lesssim z \lesssim 0.084$, but deviates at $\sim 2.9\sigma$ from the value $\Omega_{\rm HI}=(3.90\pm 0.10)\times 10^{-3}$ reported by ALFALFA at $z\sim 0$;
    \item the resulting values for $\Omega_{\rm HI}$ obtained by assuming the linear and the polynomial model for the scaling relation differ by $\sim 2.4\sigma$ in the combined stacking case.
\end{itemize}

In conclusion, we argue that a similar approach as the one in this study can be used to derive the HIMF and the $\Omega_{\rm HI}$ parameter at different redshifts, provided that a reliable scaling relation is available to convert from well-studied mass/luminosity functions (such as the ones for $M_\star$ or $M_B)$ to $M_{\rm HI}$. In particular, future efforts estimating e.g. the $M_{\rm HI}-M_\star$ scaling relation with larger samples will likely increase the number of available data points. This will help to better constrain the functional form of such a relation, and hence, to increase the accuracy of the procedure.


\begin{acknowledgements}
      The authors warmly thank Nissim Kanekar, Apurba Bera and Aditya Chowdhury for providing the measurements underlying their publications. F.S. acknowledges the support of the Swiss National Science Foundation (SNSF) 200021\_214990/1 grant. A.B. acknowledges the support of the doctoral grant funded by the University of Padova and by the Italian Ministry of Education, University and Research (MIUR). G.R. acknowledges the support from grant PRIN MIUR 2017- 20173ML3WW 001. A.B. and G.R. acknowledge support from INAF under the Large Grant 2022 funding scheme (project "MeerKAT and LOFAR Team up: a Unique Radio Window on Galaxy/AGN co-Evolution”). A.B. acknowledges financial support from the South African Department of Science and Innovation's National Research Foundation under the ISARP RADIOMAP Joint Research Scheme (DSI-NRF Grant Number 150551), and from the Italian Ministry of Foreign Affairs and International Cooperation under the “Progetti di Grande Rilevanza” scheme (project RADIOMAP, grant number ZA23GR03). M.V. acknowledges financial support from the Inter-University Institute for Data Intensive Astronomy (IDIA), a partnership of the University of Cape Town, the University of Pretoria and the University of the Western Cape, and from the South African Department of Science and Innovation's National Research Foundation under the ISARP RADIOSKY2020 and RADIOMAP Joint Research Schemes (DSI-NRF Grant Numbers 113121 and 150551) and the SRUG HIPPO Projects (DSI-NRF Grant Numbers 121291 and SRUG22031677). 
\end{acknowledgements}

%
%


\bibliographystyle{aa}
\bibliography{lit}

\begin{thebibliography}{78}
\expandafter\ifx\csname natexlab\endcsname\relax\def\natexlab#1{#1}\fi

\bibitem[{{Arnouts} \& {Ilbert}(2011)}]{Arnouts2011}
{Arnouts}, S. \& {Ilbert}, O. 2011, {LePHARE: Photometric Analysis for Redshift Estimate}, Astrophysics Source Code Library, record ascl:1108.009

\bibitem[{{Ba{\~n}ares-Hern{\'a}ndez} {et~al.}(2023){Ba{\~n}ares-Hern{\'a}ndez}, {Castillo}, {Martin Camalich}, \& {Iorio}}]{Banares2023}
{Ba{\~n}ares-Hern{\'a}ndez}, A., {Castillo}, A., {Martin Camalich}, J., \& {Iorio}, G. 2023, \aap, 676, A63

\bibitem[{{Bauer} {et~al.}(2021){Bauer}, {Marsh}, {Hlo{\v{z}}ek}, {Padmanabhan}, \& {Lagu{\"e}}}]{Bauer2021}
{Bauer}, J.~B., {Marsh}, D. J.~E., {Hlo{\v{z}}ek}, R., {Padmanabhan}, H., \& {Lagu{\"e}}, A. 2021, \mnras, 500, 3162

\bibitem[{{Beckmann} {et~al.}(2017){Beckmann}, {Devriendt}, {Slyz}, {Peirani}, {Richardson}, {Dubois}, {Pichon}, {Chisari}, {Kaviraj}, {Laigle}, \& {Volonteri}}]{Beckman2017}
{Beckmann}, R.~S., {Devriendt}, J., {Slyz}, A., {et~al.} 2017, \mnras, 472, 949

\bibitem[{{Bera} {et~al.}(2019){Bera}, {Kanekar}, {Chengalur}, \& {Bagla}}]{Bera2019}
{Bera}, A., {Kanekar}, N., {Chengalur}, J.~N., \& {Bagla}, J.~S. 2019, \apjl, 882, L7

\bibitem[{{Bera} {et~al.}(2022){Bera}, {Kanekar}, {Chengalur}, \& {Bagla}}]{Bera2022}
{Bera}, A., {Kanekar}, N., {Chengalur}, J.~N., \& {Bagla}, J.~S. 2022, \apjl, 940, L10

\bibitem[{{Berti} {et~al.}(2024){Berti}, {Spinelli}, \& {Viel}}]{Berti2024}
{Berti}, M., {Spinelli}, M., \& {Viel}, M. 2024, \mnras, 529, 4803

\bibitem[{{Bianchetti} {et~al.}(2025){Bianchetti}, {Sinigaglia}, {Rodighiero}, {Elson}, {Vaccari}, {Pisano}, {Luber}, {Prandoni}, {Hess}, {Baes}, {Adams}, {Maccagni}, {Renzini}, {Bisigello}, {Yun}, {Momjian}, {Gim}, {Pan}, {Oosterloo}, {Dodson}, {Lucero}, {Frank}, {Ilbert}, {Davies}, {Khostovan}, \& {Salvato}}]{Bianchetti2025}
{Bianchetti}, A., {Sinigaglia}, F., {Rodighiero}, G., {et~al.} 2025, \apj, 982, 82

\bibitem[{{Bigiel} {et~al.}(2008){Bigiel}, {Leroy}, {Walter}, {Brinks}, {de Blok}, {Madore}, \& {Thornley}}]{Bigiel2008}
{Bigiel}, F., {Leroy}, A., {Walter}, F., {et~al.} 2008, \aj, 136, 2846

\bibitem[{{Blitz} \& {Rosolowsky}(2006)}]{Blitz2006}
{Blitz}, L. \& {Rosolowsky}, E. 2006, \apj, 650, 933

\bibitem[{{Catinella} {et~al.}(2018){Catinella}, {Saintonge}, {Janowiecki}, {Cortese}, {Dav{\'e}}, {Lemonias}, {Cooper}, {Schiminovich}, {Hummels}, {Fabello}, {Ger{\'e}b}, {Kilborn}, \& {Wang}}]{Catinella2018}
{Catinella}, B., {Saintonge}, A., {Janowiecki}, S., {et~al.} 2018, \mnras, 476, 875

\bibitem[{{Chang} {et~al.}(2010){Chang}, {Pen}, {Bandura}, \& {Peterson}}]{Chang2010}
{Chang}, T.-C., {Pen}, U.-L., {Bandura}, K., \& {Peterson}, J.~B. 2010, \nat, 466, 463

\bibitem[{{Chen} {et~al.}(2021){Chen}, {Wolz}, {Spinelli}, \& {Murray}}]{Chen2021}
{Chen}, Z., {Wolz}, L., {Spinelli}, M., \& {Murray}, S.~G. 2021, \mnras, 502, 5259

\bibitem[{{Chowdhury} {et~al.}(2024){Chowdhury}, {Kanekar}, \& {Chengalur}}]{Chowdhury2024}
{Chowdhury}, A., {Kanekar}, N., \& {Chengalur}, J.~N. 2024, \apjl, 966, L39

\bibitem[{{Crighton} {et~al.}(2015){Crighton}, {Murphy}, {Prochaska}, {Worseck}, {Rafelski}, {Becker}, {Ellison}, {Fumagalli}, {Lopez}, {Meiksin}, \& {O'Meara}}]{Chrighton2015}
{Crighton}, N. H.~M., {Murphy}, M.~T., {Prochaska}, J.~X., {et~al.} 2015, \mnras, 452, 217

\bibitem[{{Cunnington} {et~al.}(2023){Cunnington}, {Li}, {Santos}, {Wang}, {Carucci}, {Irfan}, {Pourtsidou}, {Spinelli}, {Wolz}, {Soares}, {Blake}, {Bull}, {Engelbrecht}, {Fonseca}, {Grainge}, \& {Ma}}]{Cunnington2023}
{Cunnington}, S., {Li}, Y., {Santos}, M.~G., {et~al.} 2023, \mnras, 518, 6262

\bibitem[{{Dav{\'e}} {et~al.}(2019){Dav{\'e}}, {Angl{\'e}s-Alc{\'a}zar}, {Narayanan}, {Li}, {Rafieferantsoa}, \& {Appleby}}]{Dave2019}
{Dav{\'e}}, R., {Angl{\'e}s-Alc{\'a}zar}, D., {Narayanan}, D., {et~al.} 2019, \mnras, 486, 2827

\bibitem[{{Dav{\'e}} {et~al.}(2020){Dav{\'e}}, {Crain}, {Stevens}, {Narayanan}, {Saintonge}, {Catinella}, \& {Cortese}}]{Dave2020}
{Dav{\'e}}, R., {Crain}, R.~A., {Stevens}, A. R.~H., {et~al.} 2020, \mnras, 497, 146

\bibitem[{{Dav{\'e}} {et~al.}(2016){Dav{\'e}}, {Thompson}, \& {Hopkins}}]{Dave2016}
{Dav{\'e}}, R., {Thompson}, R., \& {Hopkins}, P.~F. 2016, \mnras, 462, 3265

\bibitem[{{Davies} {et~al.}(2018){Davies}, {Robotham}, {Driver}, {Lagos}, {Cortese}, {Mannering}, {Foster}, {Lidman}, {Hashemizadeh}, {Koushan}, {O'Toole}, {Baldry}, {Bilicki}, {Bland-Hawthorn}, {Bremer}, {Brown}, {Bryant}, {Catinella}, {Croom}, {Grootes}, {Holwerda}, {Jarvis}, {Maddox}, {Meyer}, {Moffett}, {Phillipps}, {Taylor}, {Windhorst}, \& {Wolf}}]{Davies2018}
{Davies}, L.~J.~M., {Robotham}, A.~S.~G., {Driver}, S.~P., {et~al.} 2018, \mnras, 480, 768

\bibitem[{{DeBoer} {et~al.}(2017){DeBoer}, {Parsons}, {Aguirre}, {Alexander}, {Ali}, {Beardsley}, {Bernardi}, {Bowman}, {Bradley}, {Carilli}, {Cheng}, {de Lera Acedo}, {Dillon}, {Ewall-Wice}, {Fadana}, {Fagnoni}, {Fritz}, {Furlanetto}, {Glendenning}, {Greig}, {Grobbelaar}, {Hazelton}, {Hewitt}, {Hickish}, {Jacobs}, {Julius}, {Kariseb}, {Kohn}, {Lekalake}, {Liu}, {Loots}, {MacMahon}, {Malan}, {Malgas}, {Maree}, {Martinot}, {Mathison}, {Matsetela}, {Mesinger}, {Morales}, {Neben}, {Patra}, {Pieterse}, {Pober}, {Razavi-Ghods}, {Ringuette}, {Robnett}, {Rosie}, {Sell}, {Smith}, {Syce}, {Tegmark}, {Thyagarajan}, {Williams}, \& {Zheng}}]{DeBoer2017}
{DeBoer}, D.~R., {Parsons}, A.~R., {Aguirre}, J.~E., {et~al.} 2017, \pasp, 129, 045001

\bibitem[{{Delhaize} {et~al.}(2013){Delhaize}, {Meyer}, {Staveley-Smith}, \& {Boyle}}]{Delhaize2013}
{Delhaize}, J., {Meyer}, M.~J., {Staveley-Smith}, L., \& {Boyle}, B.~J. 2013, \mnras, 433, 1398

\bibitem[{{DESI Collaboration} {et~al.}(2024){DESI Collaboration}, {Adame}, {Aguilar}, {Ahlen}, {Alam}, {Aldering}, {Alexander}, {Alfarsy}, {Allende Prieto}, {Alvarez}, {Alves}, {Anand}, {Andrade-Oliveira}, {Armengaud}, {Asorey}, {Avila}, {Aviles}, {Bailey}, {Balaguera-Antol{\'\i}nez}, {Ballester}, {Baltay}, {Bault}, {Bautista}, {Behera}, {Beltran}, {BenZvi}, {Beraldo e Silva}, {Bermejo-Climent}, {Berti}, {Besuner}, {Beutler}, {Bianchi}, {Blake}, {Blum}, {Bolton}, {Brieden}, {Brodzeller}, {Brooks}, {Brown}, {Buckley-Geer}, {Burtin}, {Cabayol-Garcia}, {Cai}, {Canning}, {Cardiel-Sas}, {Carnero Rosell}, {Castander}, {Cervantes-Cota}, {Chabanier}, {Chaussidon}, {Chaves-Montero}, {Chen}, {Chen}, {Chuang}, {Claybaugh}, {Cole}, {Cooper}, {Cuceu}, {Davis}, {Dawson}, {de Belsunce}, {de la Cruz}, {de la Macorra}, {Della Costa}, {de Mattia}, {Demina}, {Demirbozan}, {DeRose}, {Dey}, {Dey}, {Dhungana}, {Ding}, {Ding}, {Doel}, {Doshi}, {Douglass}, {Edge}, {Eftekharzadeh}, {Eisenstein}, {Elliott}, {Ereza}, {Escoffier},
  {Fagrelius}, {Fan}, {Fanning}, {Fawcett}, {Ferraro}, {Flaugher}, {Font-Ribera}, {Forero-Romero}, {Forero-S{\'a}nchez}, {Frenk}, {G{\"a}nsicke}, {Garc{\'\i}a}, {Garc{\'\i}a-Bellido}, {Garcia-Quintero}, {Garrison}, {Gil-Mar{\'\i}n}, {Golden-Marx}, {Gontcho A Gontcho}, {Gonzalez-Morales}, {Gonzalez-Perez}, {Gordon}, {Graur}, {Green}, {Gruen}, {Guy}, {Hadzhiyska}, {Hahn}, {Han}, {Hanif}, {Herrera-Alcantar}, {Honscheid}, {Hou}, {Howlett}, {Huterer}, {Ir{\v{s}}i{\v{c}}}, {Ishak}, {Jacques}, {Jana}, {Jiang}, {Jimenez}, {Jing}, {Joudaki}, {Joyce}, {Jullo}, {Juneau}, {Kara{\c{c}}ayl{\i}}, {Karim}, {Kehoe}, {Kent}, {Khederlarian}, {Kim}, {Kirkby}, {Kisner}, {Kitaura}, {Kizhuprakkat}, {Kneib}, {Koposov}, {Kov{\'a}cs}, {Kremin}, {Krolewski}, {L'Huillier}, {Lahav}, {Lambert}, {Lamman}, {Lan}, {Landriau}, {Lang}, {Lange}, {Lasker}, {Leauthaud}, {Le Guillou}, {Levi}, {Li}, {Linder}, {Lyons}, {Magneville}, {Manera}, {Manser}, {Margala}, {Martini}, {McDonald}, {Medina}, {Medina-Varela}, {Meisner}, {Mena-Fern{\'a}ndez},
  {Meneses-Rizo}, {Mezcua}, {Miquel}, {Montero-Camacho}, {Moon}, {Moore}, {Moustakas}, {Mueller}, {Mundet}, {Mu{\~n}oz-Guti{\'e}rrez}, {Myers}, {Nadathur}, {Napolitano}, {Neveux}, {Newman}, {Nie}, {Nikutta}, {Niz}, {Norberg}, {Noriega}, {Paillas}, {Palanque-Delabrouille}, {Palmese}, {Pan}, {Parkinson}, {Penmetsa}, {Percival}, {P{\'e}rez-Fern{\'a}ndez}, {P{\'e}rez-R{\`a}fols}, {Pieri}, {Poppett}, {Porredon}, \& {Pothier}}]{DESI2023}
{DESI Collaboration}, {Adame}, A.~G., {Aguilar}, J., {et~al.} 2024, \aj, 168, 58

\bibitem[{{Diemer} {et~al.}(2018){Diemer}, {Stevens}, {Forbes}, {Marinacci}, {Hernquist}, {Lagos}, {Sternberg}, {Pillepich}, {Nelson}, {Popping}, {Villaescusa-Navarro}, {Torrey}, \& {Vogelsberger}}]{Diemer2018}
{Diemer}, B., {Stevens}, A. R.~H., {Forbes}, J.~C., {et~al.} 2018, \apjs, 238, 33

\bibitem[{{Durbala} {et~al.}(2020){Durbala}, {Finn}, {Crone Odekon}, {Haynes}, {Koopmann}, \& {O'Donoghue}}]{Durbala2020}
{Durbala}, A., {Finn}, R.~A., {Crone Odekon}, M., {et~al.} 2020, \aj, 160, 271

\bibitem[{{Feldmann} {et~al.}(2023){Feldmann}, {Quataert}, {Faucher-Gigu{\`e}re}, {Hopkins}, {{\c{C}}atmabacak}, {Kere{\v{s}}}, {Bassini}, {Bernardini}, {Bullock}, {Cenci}, {Gensior}, {Liang}, {Moreno}, \& {Wetzel}}]{Feldmann2023}
{Feldmann}, R., {Quataert}, E., {Faucher-Gigu{\`e}re}, C.-A., {et~al.} 2023, \mnras, 522, 3831

\bibitem[{Fernández {et~al.}(2013)Fernández, van Gorkom, Hess, Pisano, Kreckel, Momjian, Popping, Oosterloo, Chomiuk, Verheijen, Henning, Schiminovich, Bershady, Wilcots, \& Scoville}]{Fernandez2013}
Fernández, X., van Gorkom, J.~H., Hess, K.~M., {et~al.} 2013, The Astrophysical Journal Letters, 770, L29

\bibitem[{{Foreman-Mackey} {et~al.}(2013){Foreman-Mackey}, {Hogg}, {Lang}, \& {Goodman}}]{ForemanMackey2013}
{Foreman-Mackey}, D., {Hogg}, D.~W., {Lang}, D., \& {Goodman}, J. 2013, \pasp, 125, 306

\bibitem[{{Gal{\'a}rraga-Espinosa} {et~al.}(2021){Gal{\'a}rraga-Espinosa}, {Aghanim}, {Langer}, \& {Tanimura}}]{Galarraga2021}
{Gal{\'a}rraga-Espinosa}, D., {Aghanim}, N., {Langer}, M., \& {Tanimura}, H. 2021, \aap, 649, A117

\bibitem[{{Gensior} {et~al.}(2023){Gensior}, {Feldmann}, {Mayer}, {Wetzel}, {Hopkins}, \& {Faucher-Gigu{\`e}re}}]{Gensior2023}
{Gensior}, J., {Feldmann}, R., {Mayer}, L., {et~al.} 2023, \mnras, 518, L63

\bibitem[{{Gensior} {et~al.}(2024){Gensior}, {Feldmann}, {Reina-Campos}, {Trujillo-Gomez}, {Mayer}, {Keller}, {Wetzel}, {Kruijssen}, {Hopkins}, \& {Moreno}}]{Gensior2024}
{Gensior}, J., {Feldmann}, R., {Reina-Campos}, M., {et~al.} 2024, \mnras, 531, 1158

\bibitem[{{Gnedin} \& {Kravtsov}(2011)}]{GnedinKravtsov2011}
{Gnedin}, N.~Y. \& {Kravtsov}, A.~V. 2011, \apj, 728, 88

\bibitem[{Gogate(2022)}]{Gogate2022}
Gogate, A. 2022, PhD thesis, University of Groningen

\bibitem[{{Hoppmann} {et~al.}(2015){Hoppmann}, {Staveley-Smith}, {Freudling}, {Zwaan}, {Minchin}, \& {Calabretta}}]{Hoppman2015}
{Hoppmann}, L., {Staveley-Smith}, L., {Freudling}, W., {et~al.} 2015, \mnras, 452, 3726

\bibitem[{{Hu} {et~al.}(2019){Hu}, {Hoppmann}, {Staveley-Smith}, {Ger{\'e}b}, {Oosterloo}, {Morganti}, {Catinella}, {Cortese}, {Lagos}, \& {Meyer}}]{Hu2019}
{Hu}, W., {Hoppmann}, L., {Staveley-Smith}, L., {et~al.} 2019, \mnras, 489, 1619

\bibitem[{{Huang} {et~al.}(2012){Huang}, {Haynes}, {Giovanelli}, \& {Brinchmann}}]{Huang2012}
{Huang}, S., {Haynes}, M.~P., {Giovanelli}, R., \& {Brinchmann}, J. 2012, \apj, 756, 113

\bibitem[{{Ilbert} {et~al.}(2006){Ilbert}, {Arnouts}, {McCracken}, {Bolzonella}, {Bertin}, {Le F{\`e}vre}, {Mellier}, {Zamorani}, {Pell{\`o}}, {Iovino}, {Tresse}, {Le Brun}, {Bottini}, {Garilli}, {Maccagni}, {Picat}, {Scaramella}, {Scodeggio}, {Vettolani}, {Zanichelli}, {Adami}, {Bardelli}, {Cappi}, {Charlot}, {Ciliegi}, {Contini}, {Cucciati}, {Foucaud}, {Franzetti}, {Gavignaud}, {Guzzo}, {Marano}, {Marinoni}, {Mazure}, {Meneux}, {Merighi}, {Paltani}, {Pollo}, {Pozzetti}, {Radovich}, {Zucca}, {Bondi}, {Bongiorno}, {Busarello}, {de La Torre}, {Gregorini}, {Lamareille}, {Mathez}, {Merluzzi}, {Ripepi}, {Rizzo}, \& {Vergani}}]{Ilbert2006}
{Ilbert}, O., {Arnouts}, S., {McCracken}, H.~J., {et~al.} 2006, \aap, 457, 841

\bibitem[{{Jarvis} {et~al.}(2016){Jarvis}, {Taylor}, {Agudo}, {Allison}, {Deane}, {Frank}, {Gupta}, {Heywood}, {Maddox}, {McAlpine}, {Santos}, {Scaife}, {Vaccari}, {Zwart}, {Adams}, {Bacon}, {Baker}, {Bassett}, {Best}, {Beswick}, {Blyth}, {Brown}, {Bruggen}, {Cluver}, {Colafrancesco}, {Cotter}, {Cress}, {Dav{\'e}}, {Ferrari}, {Hardcastle}, {Hale}, {Harrison}, {Hatfield}, {Klockner}, {Kolwa}, {Malefahlo}, {Marubini}, {Mauch}, {Moodley}, {Morganti}, {Norris}, {Peters}, {Prandoni}, {Prescott}, {Oliver}, {Oozeer}, {Rottgering}, {Seymour}, {Simpson}, {Smirnov}, \& {Smith}}]{Jarvis2016}
{Jarvis}, M., {Taylor}, R., {Agudo}, I., {et~al.} 2016, in MeerKAT Science: On the Pathway to the SKA, 6

\bibitem[{{Jonas} \& {MeerKAT Team}(2016)}]{Jonas2016}
{Jonas}, J. \& {MeerKAT Team}. 2016, in MeerKAT Science: On the Pathway to the SKA, 1

\bibitem[{{Jones} {et~al.}(2018){Jones}, {Haynes}, {Giovanelli}, \& {Moorman}}]{Jones2018}
{Jones}, M.~G., {Haynes}, M.~P., {Giovanelli}, R., \& {Moorman}, C. 2018, \mnras, 477, 2

\bibitem[{{Khostovan} {et~al.}(2025){Khostovan}, {Kartaltepe}, {Salvato}, {Ilbert}, {Casey}, {Algera}, {Antwi-Danso}, {Battisti}, {Brinch}, {Brusa}, {Calabro}, {Capak}, {Chartab}, {Cooper}, {Cox}, {Darvish}, {Drakos}, {Faisst}, {George}, {Gozaliasl}, {Harish}, {Hasinger}, {Hatamnia}, {Iovino}, {Jin}, {Kashino}, {Koekemoer}, {Laishram}, {Lee}, {Lertprasertpong}, {Lilly}, {Masters}, {Mobasher}, {Nagao}, {Onodera}, {Peng}, {Sanders}, {Sanders}, {Sattari}, {Scoville}, {Shah}, {Silverman}, {Suzuki}, {Tanaka}, {Toft}, {Trakhtenbrot}, {Trump}, {Vaccari}, {Valentino}, {Vanderhoof}, {Weaver}, {Yun}, \& {Zavala}}]{Khostovan2025}
{Khostovan}, A.~A., {Kartaltepe}, J.~S., {Salvato}, M., {et~al.} 2025, arXiv e-prints, arXiv:2503.00120

\bibitem[{{Koopmans} {et~al.}(2015){Koopmans}, {Pritchard}, {Mellema}, {Aguirre}, {Ahn}, {Barkana}, {van Bemmel}, {Bernardi}, {Bonaldi}, {Briggs}, {de Bruyn}, {Chang}, {Chapman}, {Chen}, {Ciardi}, {Dayal}, {Ferrara}, {Fialkov}, {Fiore}, {Ichiki}, {Illiev}, {Inoue}, {Jelic}, {Jones}, {Lazio}, {Maio}, {Majumdar}, {Mack}, {Mesinger}, {Morales}, {Parsons}, {Pen}, {Santos}, {Schneider}, {Semelin}, {de Souza}, {Subrahmanyan}, {Takeuchi}, {Vedantham}, {Wagg}, {Webster}, {Wyithe}, {Datta}, \& {Trott}}]{Koopman2015}
{Koopmans}, L., {Pritchard}, J., {Mellema}, G., {et~al.} 2015, in Advancing Astrophysics with the Square Kilometre Array (AASKA14), 1

\bibitem[{{Krumholz} \& {Gnedin}(2011)}]{KrumholzGnedin2011}
{Krumholz}, M.~R. \& {Gnedin}, N.~Y. 2011, \apj, 729, 36

\bibitem[{{Lah} {et~al.}(2007){Lah}, {Chengalur}, {Briggs}, {Colless}, {de Propris}, {Pracy}, {de Blok}, {Fujita}, {Ajiki}, {Shioya}, {Nagao}, {Murayama}, {Taniguchi}, {Yagi}, \& {Okamura}}]{Lah2007}
{Lah}, P., {Chengalur}, J.~N., {Briggs}, F.~H., {et~al.} 2007, \mnras, 376, 1357

\bibitem[{{Maddox} {et~al.}(2021){Maddox}, {Frank}, {Ponomareva}, {Jarvis}, {Adams}, {Dav{\'e}}, {Oosterloo}, {Santos}, {Blyth}, {Glowacki}, {Kraan-Korteweg}, {Mulaudzi}, {Namumba}, {Prandoni}, {Rajohnson}, {Spekkens}, {Adams}, {Bowler}, {Collier}, {Heywood}, {Sekhar}, \& {Taylor}}]{Maddox2021}
{Maddox}, N., {Frank}, B.~S., {Ponomareva}, A.~A., {et~al.} 2021, \aap, 646, A35

\bibitem[{{Maddox} {et~al.}(2015){Maddox}, {Hess}, {Obreschkow}, {Jarvis}, \& {Blyth}}]{Maddox2015}
{Maddox}, N., {Hess}, K.~M., {Obreschkow}, D., {Jarvis}, M.~J., \& {Blyth}, S.~L. 2015, \mnras, 447, 1610

\bibitem[{{Martin} {et~al.}(2010){Martin}, {Papastergis}, {Giovanelli}, {Haynes}, {Springob}, \& {Stierwalt}}]{Martin2010}
{Martin}, A.~M., {Papastergis}, E., {Giovanelli}, R., {et~al.} 2010, \apj, 723, 1359

\bibitem[{{Martizzi} {et~al.}(2019){Martizzi}, {Vogelsberger}, {Artale}, {Haider}, {Torrey}, {Marinacci}, {Nelson}, {Pillepich}, {Weinberger}, {Hernquist}, {Naiman}, \& {Springel}}]{Martizzi2019}
{Martizzi}, D., {Vogelsberger}, M., {Artale}, M.~C., {et~al.} 2019, \mnras, 486, 3766

\bibitem[{{Masui} {et~al.}(2013){Masui}, {Switzer}, {Banavar}, {Bandura}, {Blake}, {Calin}, {Chang}, {Chen}, {Li}, {Liao}, {Natarajan}, {Pen}, {Peterson}, {Shaw}, \& {Voytek}}]{Masui2013}
{Masui}, K.~W., {Switzer}, E.~R., {Banavar}, N., {et~al.} 2013, \apjl, 763, L20

\bibitem[{{Nelson} {et~al.}(2015){Nelson}, {Pillepich}, {Genel}, {Vogelsberger}, {Springel}, {Torrey}, {Rodriguez-Gomez}, {Sijacki}, {Snyder}, {Griffen}, {Marinacci}, {Blecha}, {Sales}, {Xu}, \& {Hernquist}}]{Nelson2015}
{Nelson}, D., {Pillepich}, A., {Genel}, S., {et~al.} 2015, Astronomy and Computing, 13, 12

\bibitem[{{Nelson} {et~al.}(2019){Nelson}, {Springel}, {Pillepich}, {Rodriguez-Gomez}, {Torrey}, {Genel}, {Vogelsberger}, {Pakmor}, {Marinacci}, {Weinberger}, {Kelley}, {Lovell}, {Diemer}, \& {Hernquist}}]{Nelson2019}
{Nelson}, D., {Springel}, V., {Pillepich}, A., {et~al.} 2019, Computational Astrophysics and Cosmology, 6, 2

\bibitem[{{Noterdaeme} {et~al.}(2012){Noterdaeme}, {Petitjean}, {Carithers}, {P{\^a}ris}, {Font-Ribera}, {Bailey}, {Aubourg}, {Bizyaev}, {Ebelke}, {Finley}, {Ge}, {Malanushenko}, {Malanushenko}, {Miralda-Escud{\'e}}, {Myers}, {Oravetz}, {Pan}, {Pieri}, {Ross}, {Schneider}, {Simmons}, \& {York}}]{Noterdaeme2012}
{Noterdaeme}, P., {Petitjean}, P., {Carithers}, W.~C., {et~al.} 2012, \aap, 547, L1

\bibitem[{{O'Brien} {et~al.}(2018){O'Brien}, {Chiarelli}, {Dentico}, {Stulge}, {Stefanski}, {Moss}, \& {Chaykov}}]{OBrien2018}
{O'Brien}, J.~G., {Chiarelli}, T.~L., {Dentico}, J., {et~al.} 2018, \apj, 852, 6

\bibitem[{{Pan} {et~al.}(2023){Pan}, {Jarvis}, {Santos}, {Maddox}, {Frank}, {Ponomareva}, {Prandoni}, {Kurapati}, {Baes}, {Mancera Pi{\~n}a}, {Rodighiero}, {Meyer}, {Dav{\'e}}, {Sharma}, {Rajohnson}, {Adams}, {Bowler}, {Sinigaglia}, {van der Hulst}, {Hatfield}, {Sekhar}, \& {Collier}}]{Pan2023}
{Pan}, H., {Jarvis}, M.~J., {Santos}, M.~G., {et~al.} 2023, \mnras, 525, 256

\bibitem[{{Parkash} {et~al.}(2018){Parkash}, {Brown}, {Jarrett}, \& {Bonne}}]{Parkash2018}
{Parkash}, V., {Brown}, M. J.~I., {Jarrett}, T.~H., \& {Bonne}, N.~J. 2018, \apj, 864, 40

\bibitem[{{Paul} {et~al.}(2023){Paul}, {Santos}, {Chen}, \& {Wolz}}]{Paul2023}
{Paul}, S., {Santos}, M.~G., {Chen}, Z., \& {Wolz}, L. 2023, arXiv e-prints, arXiv:2301.11943

\bibitem[{{Pinetti} {et~al.}(2020){Pinetti}, {Camera}, {Fornengo}, \& {Regis}}]{Pinetti2020}
{Pinetti}, E., {Camera}, S., {Fornengo}, N., \& {Regis}, M. 2020, \jcap, 2020, 044

\bibitem[{{Ponomareva} {et~al.}(2023){Ponomareva}, {Jarvis}, {Pan}, {Maddox}, {Jones}, {Frank}, {Rajohnson}, {Mulaudzi}, {Meyer}, {Adams}, {Baes}, {Hess}, {Kurapati}, {Prandoni}, {Sinigaglia}, {Spekkens}, {Tudorache}, {Heywood}, {Collier}, \& {Sekhar}}]{Ponomareva2023}
{Ponomareva}, A.~A., {Jarvis}, M.~J., {Pan}, H., {et~al.} 2023, \mnras, 522, 5308

\bibitem[{{Primack}(2024)}]{Primack2024}
{Primack}, J.~R. 2024, Annual Review of Nuclear and Particle Science, 74, 173

\bibitem[{{Rao} {et~al.}(2006){Rao}, {Turnshek}, \& {Nestor}}]{Rao2006}
{Rao}, S.~M., {Turnshek}, D.~A., \& {Nestor}, D.~B. 2006, \apj, 636, 610

\bibitem[{{Rao} {et~al.}(2017){Rao}, {Turnshek}, {Sardane}, \& {Monier}}]{Rao2017}
{Rao}, S.~M., {Turnshek}, D.~A., {Sardane}, G.~M., \& {Monier}, E.~M. 2017, \mnras, 471, 3428

\bibitem[{{Reina-Campos} {et~al.}(2022){Reina-Campos}, {Keller}, {Kruijssen}, {Gensior}, {Trujillo-Gomez}, {Jeffreson}, {Pfeffer}, \& {Sills}}]{ReinaCampos2022}
{Reina-Campos}, M., {Keller}, B.~W., {Kruijssen}, J.~M.~D., {et~al.} 2022, \mnras, 517, 3144

\bibitem[{{Rhee} {et~al.}(2018){Rhee}, {Lah}, {Briggs}, {Chengalur}, {Colless}, {Willner}, {Ashby}, \& {Le F{\`e}vre}}]{Rhee2018}
{Rhee}, J., {Lah}, P., {Briggs}, F.~H., {et~al.} 2018, \mnras, 473, 1879

\bibitem[{{Rhee} {et~al.}(2016){Rhee}, {Lah}, {Chengalur}, {Briggs}, \& {Colless}}]{Rhee2016}
{Rhee}, J., {Lah}, P., {Chengalur}, J.~N., {Briggs}, F.~H., \& {Colless}, M. 2016, \mnras, 460, 2675

\bibitem[{{Santos} {et~al.}(2015){Santos}, {Bull}, {Alonso}, {Camera}, {Ferreira}, {Bernardi}, {Maartens}, {Viel}, {Villaescusa-Navarro}, {Abdalla}, {Jarvis}, {Metcalf}, {Pourtsidou}, \& {Wolz}}]{Santos2015}
{Santos}, M., {Bull}, P., {Alonso}, D., {et~al.} 2015, in Advancing Astrophysics with the Square Kilometre Array (AASKA14), 19

\bibitem[{{Scoville} {et~al.}(2007){Scoville}, {Aussel}, {Brusa}, {Capak}, {Carollo}, {Elvis}, {Giavalisco}, {Guzzo}, {Hasinger}, {Impey}, {Kneib}, {LeFevre}, {Lilly}, {Mobasher}, {Renzini}, {Rich}, {Sanders}, {Schinnerer}, {Schminovich}, {Shopbell}, {Taniguchi}, \& {Tyson}}]{Scoville2007}
{Scoville}, N., {Aussel}, H., {Brusa}, M., {et~al.} 2007, \apjs, 172, 1

\bibitem[{{Sinigaglia} {et~al.}(2021){Sinigaglia}, {Kitaura}, {Balaguera-Antol{\'\i}nez}, {Nagamine}, {Ata}, {Shimizu}, \& {S{\'a}nchez-Benavente}}]{Sinigaglia2021}
{Sinigaglia}, F., {Kitaura}, F.-S., {Balaguera-Antol{\'\i}nez}, A., {et~al.} 2021, \apj, 921, 66

\bibitem[{{Sinigaglia} {et~al.}(2024){Sinigaglia}, {Rodighiero}, {Elson}, {Bianchetti}, {Vaccari}, {Maddox}, {Ponomareva}, {Frank}, {Jarvis}, {Catinella}, {Cortese}, {Roychowdhury}, {Baes}, {Collier}, {Ilbert}, {Khostovan}, {Kurapati}, {Pan}, {Prandoni}, {Rajohnson}, {Salvato}, {Sekhar}, \& {Sharma}}]{Sinigaglia2024}
{Sinigaglia}, F., {Rodighiero}, G., {Elson}, E., {et~al.} 2024, \mnras, 529, 4192

\bibitem[{{Sinigaglia} {et~al.}(2022){Sinigaglia}, {Rodighiero}, {Elson}, {Vaccari}, {Maddox}, {Frank}, {Jarvis}, {Oosterloo}, {Dav{\'e}}, {Salvato}, {Baes}, {Bellstedt}, {Bisigello}, {Collier}, {Cook}, {Davies}, {Delhaize}, {Driver}, {Foster}, {Kurapati}, {Lagos}, {Lidman}, {Mancera Pi{\~n}a}, {Meyer}, {Mogotsi}, {Pan}, {Ponomareva}, {Prandoni}, {Rajohnson}, {Robotham}, {Santos}, {Sekhar}, {Spekkens}, {Thorne}, {van der Hulst}, \& {Wong}}]{Sinigaglia2022}
{Sinigaglia}, F., {Rodighiero}, G., {Elson}, E., {et~al.} 2022, \apjl, 935, L13

\bibitem[{{Somerville} \& {Dav{\'e}}(2015)}]{Somerville2015}
{Somerville}, R.~S. \& {Dav{\'e}}, R. 2015, \araa, 53, 51

\bibitem[{{Springel}(2010)}]{Springel2010}
{Springel}, V. 2010, \mnras, 401, 791

\bibitem[{{Sternberg} {et~al.}(2014){Sternberg}, {Le Petit}, {Roueff}, \& {Le Bourlot}}]{Sternberg2014}
{Sternberg}, A., {Le Petit}, F., {Roueff}, E., \& {Le Bourlot}, J. 2014, \apj, 790, 10

\bibitem[{{Vogelsberger} {et~al.}(2020){Vogelsberger}, {Marinacci}, {Torrey}, \& {Puchwein}}]{Vogelsberger2020}
{Vogelsberger}, M., {Marinacci}, F., {Torrey}, P., \& {Puchwein}, E. 2020, Nature Reviews Physics, 2, 42

\bibitem[{{Weaver} {et~al.}(2023){Weaver}, {Davidzon}, {Toft}, {Ilbert}, {McCracken}, {Gould}, {Jespersen}, {Steinhardt}, {Lagos}, {Capak}, {Casey}, {Chartab}, {Faisst}, {Hayward}, {Kartaltepe}, {Kauffmann}, {Koekemoer}, {Kokorev}, {Laigle}, {Liu}, {Long}, {Magdis}, {McPartland}, {Milvang-Jensen}, {Mobasher}, {Moneti}, {Peng}, {Sanders}, {Shuntov}, {Sneppen}, {Valentino}, {Zalesky}, \& {Zamorani}}]{Weaver2023}
{Weaver}, J.~R., {Davidzon}, I., {Toft}, S., {et~al.} 2023, \aap, 677, A184

\bibitem[{{Weaver} {et~al.}(2022){Weaver}, {Kauffmann}, {Ilbert}, {McCracken}, {Moneti}, {Toft}, {Brammer}, {Shuntov}, {Davidzon}, {Hsieh}, {Laigle}, {Anastasiou}, {Jespersen}, {Vinther}, {Capak}, {Casey}, {McPartland}, {Milvang-Jensen}, {Mobasher}, {Sanders}, {Zalesky}, {Arnouts}, {Aussel}, {Dunlop}, {Faisst}, {Franx}, {Furtak}, {Fynbo}, {Gould}, {Greve}, {Gwyn}, {Kartaltepe}, {Kashino}, {Koekemoer}, {Kokorev}, {Le F{\`e}vre}, {Lilly}, {Masters}, {Magdis}, {Mehta}, {Peng}, {Riechers}, {Salvato}, {Sawicki}, {Scarlata}, {Scoville}, {Shirley}, {Silverman}, {Sneppen}, {Smolc̆i{\'c}}, {Steinhardt}, {Stern}, {Tanaka}, {Taniguchi}, {Teplitz}, {Vaccari}, {Wang}, \& {Zamorani}}]{Weaver2022}
{Weaver}, J.~R., {Kauffmann}, O.~B., {Ilbert}, O., {et~al.} 2022, \apjs, 258, 11

\bibitem[{{Wolz} {et~al.}(2022){Wolz}, {Pourtsidou}, {Masui}, {Chang}, {Bautista}, {M{\"u}ller}, {Avila}, {Bacon}, {Percival}, {Cunnington}, {Anderson}, {Chen}, {Kneib}, {Li}, {Liao}, {Pen}, {Peterson}, {Rossi}, {Schneider}, {Yadav}, \& {Zhao}}]{Wolz2022}
{Wolz}, L., {Pourtsidou}, A., {Masui}, K.~W., {et~al.} 2022, \mnras, 510, 3495

\bibitem[{{Zhang} {et~al.}(2023){Zhang}, {Li}, {Zhang}, \& {Zhang}}]{Zhang2023}
{Zhang}, M., {Li}, Y., {Zhang}, J.-F., \& {Zhang}, X. 2023, \mnras, 524, 2420

\bibitem[{{Zwaan} {et~al.}(2003){Zwaan}, {Staveley-Smith}, {Koribalski}, {Henning}, {Kilborn}, {Ryder}, {Barnes}, {Bhathal}, {Boyce}, {de Blok}, {Disney}, {Drinkwater}, {Ekers}, {Freeman}, {Gibson}, {Green}, {Haynes}, {Jerjen}, {Juraszek}, {Kesteven}, {Knezek}, {Kraan-Korteweg}, {Mader}, {Marquarding}, {Meyer}, {Minchin}, {Mould}, {O'Brien}, {Oosterloo}, {Price}, {Putman}, {Ryan-Weber}, {Sadler}, {Schr{\"o}der}, {Stewart}, {Stootman}, {Warren}, {Waugh}, {Webster}, \& {Wright}}]{Zwaan2003}
{Zwaan}, M.~A., {Staveley-Smith}, L., {Koribalski}, B.~S., {et~al.} 2003, \aj, 125, 2842

\end{thebibliography}

\begin{appendix}

\section{Consistency of the HIMF from MIGHTEE and CHILES}\label{app:survey_consistency}

\begin{figure}
    \centering
    \includegraphics[width=\columnwidth]{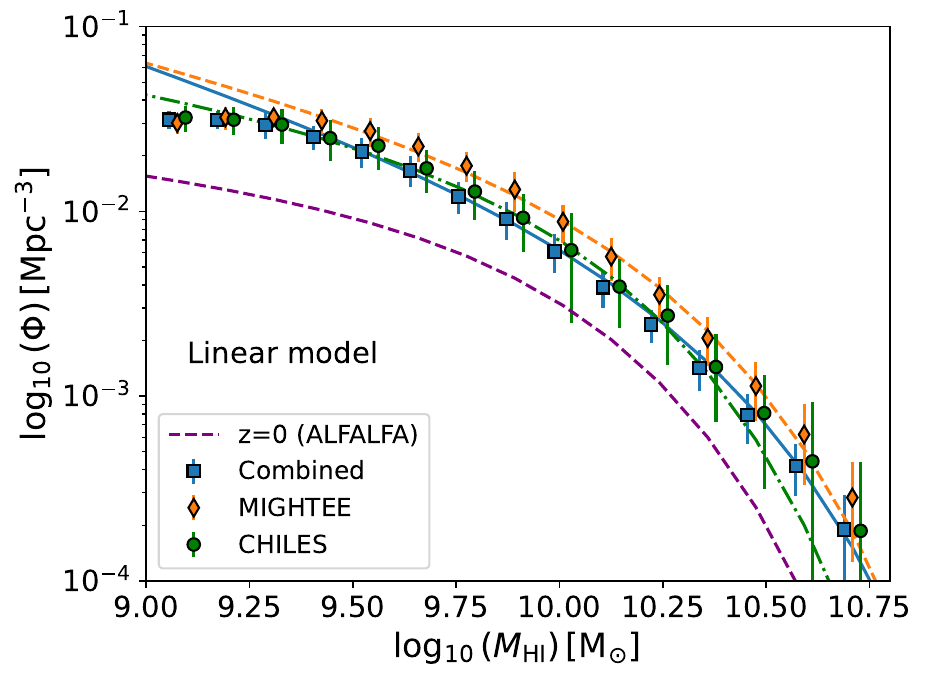}
    \includegraphics[width=\columnwidth]{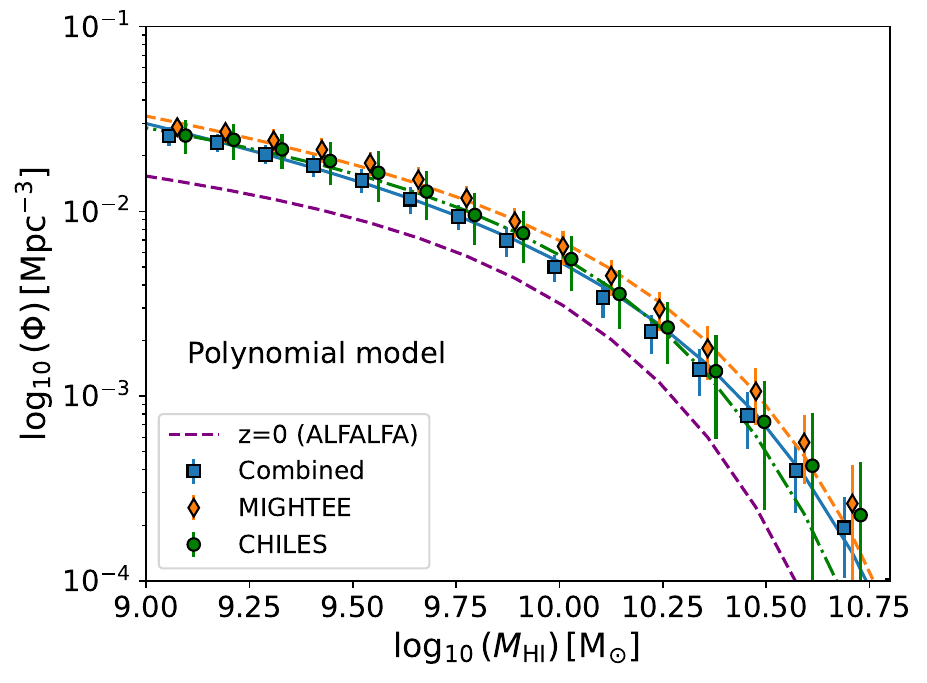}
    \caption{HIMF from the combined stacking, MIGHTEE stacking, and CHILES stacking. Top: HIMF results adopting the linear model for the $M_{\rm HI}-M_\star$ scaling relation. Bottom: HIMF results adopting the polynomial model for the $M_{\rm HI}-M_\star$ scaling relation. We display the HIMF from the combined stacking, MIGHTEE stacking, and CHILES stacking as blue squares, orange diamonds, and green circles respectively. We also show the best-fitting Schechter models as blue solid, orange dashed, and green dotted-dhased lines, respectvely. We display the ALFALFA HIMF at $z\sim 0$ as a purple dashed line for comparison.}
    \label{fig:himf_surveys}
\end{figure}

\begin{table}[]
    \centering
    \footnotesize
    \begin{tabular}{lccccc}
    \toprule
       Survey   & $\log_{10}(\tilde{M}/{\rm M}_{\odot})$ & $\alpha$ & $\log_{10}(\tilde{\Phi}/{\rm Mpc}^{-3})$ & 
       $\Omega_{\rm HI}\times 10^{-4}$ \\
       \midrule
       & & Poly. model & & \\
       \midrule
       MIGHTEE &  $10.13^{+0.08}_{-0.09}$ & $-1.38^{+0.11}_{-0.10}$ & $-2.24^{+0.13}_{-0.13}$ & $8.16^{+0.59}_{-0.54}$ \\
       \addlinespace[1.5mm]
       CHILES  &  $10.01^{+0.16}_{-0.15}$ & $-1.31^{+0.23}_{-0.20}$ & $-2.18^{+0.21}_{-0.25}$ & $6.65^{+0.99}_{-0.69}$ \\
       \addlinespace[1.5mm]
       Combined &  $10.16^{+0.08}_{-0.08}$ & $-1.48^{+0.10}_{-0.09}$ & $-2.41^{+0.12}_{-0.13}$ & $7.02^{+0.59}_{-0.52}$ \\
       \midrule
       & & Lin. model & & \\
       \midrule
       MIGHTEE &  $10.01^{+0.10}_{-0.09}$ & $-1.27^{+0.17}_{-0.16}$ & $-1.96^{+0.14}_{-0.16}$ & $1.04^{+0.94}_{-0.79}$ \\
       \addlinespace[1.5mm]
       CHILES  &  $9.95^{+0.17}_{-0.16}$ & $-1.30^{+0.28}_{-0.25}$ & $-2.01^{+0.22}_{-0.27}$ & $8.49^{+1.70}_{-1.16}$ \\
       \addlinespace[1.5mm]
       Combined & $10.13^{+0.09}_{-0.09}$ & $-1.56^{+0.14}_{-0.13}$ & $-2.31^{+0.15}_{-0.17}$ & $9.98^{+0.11}_{-0.18}$ \\
        \bottomrule
    \end{tabular}
    \vspace{1mm}
    \caption{Numerical results for the best-fitting parameters to the HIMF using a Schechter model as in Eq. \ref{eq:single_schechter}, for the results from the MIGHTEE, CHILES, and combined stacking.}
    \label{tab:results_himf_consistency}
\end{table}

In Sect. \ref{sec:results:himf}, we present the results obtained by assuming as fiducial case the combined stacking. Here, we test the consistency between such results and the ones separately from the MIGHTEE and CHILEs surveys. Fig. \ref{fig:himf_surveys} shows the resulting HIMF for the combined (blue squares), MIGHTEE (orange diamonds), and CHILES (green circles) for the linear (top panel) and polynomial (bottom panel) models
for the $M_{\rm HI}-M_\star$ relation. In addition, we display the best fit Schechter profile for the combined (blue solid), MIGHTEE (orange dashed), and CHILES (green dotted-dashed), as well as the ALFALFA HIMF at $z\sim 0$. We notice that the results from the three cases are always consistent within $1\sigma$ uncertainties. In addition, the uncertainties scale with the sample size, the combined case having the smallest and the CHILES case having the largest. This fact reflects that we are effectively incorporating the limited sample size in our framework and that taking it into account is an important piece of our modelling. 

As can be observed numerically from Table \ref{tab:results_himf_consistency}, both for the linear and the polynomial models, MIGHTEE and CHILES predict the largest and the smallest values for $\Omega_{\rm HI}$ among the ones derived in this work, while the combined case yields an intermediate value for $\Omega_{\rm HI}$ between the other two, as expected from averaging the results from the two surveys. In addition, the MIGHTEE result obtained by using the polynomial model at $z\sim 0.37$ $\Omega_{\rm HI}=8.16^{+0.59}_{-0.54}\times 10^{-3}$ is larger than the values reported by MIGHTEE at $0\lesssim z \lesssim 0.084$ $\Omega_{\rm HI}=5.26^{+0.91}_{-0.95}\times 10^{-3}$ ($1/V_{\rm max}$ method) by $\sim 2.8\sigma$ and $\Omega_{\rm HI}=6.08^{+0.30}_{-0.30}\times 10^{-3}$ (MML method) by $\sim 3.3\sigma$;

The best-fitting values for the parameters of the Schechter model are always in agreement within $1\sigma$ in the combined, MIGHTEE, and CHILES cases. When comparing the results from the linear and the polynomial model, we report a systematic shift of $\log_{10}(\tilde{M}/{\rm M}_{\odot})$ towards larger values and of $\alpha$ and $\log_{10}(\tilde{\Phi}/{\rm Mpc}^{-3})$ towards smaller values when passing from the linear to the polynomial model. While these shifts are not statistically significant and are often $<1\sigma$, we observe them systematically for all the three cases. In addition, we show in Appendix \ref{app:posteriors} the posterior distributions for the different cases. Figs. \ref{fig:posteriors_linear} and \ref{fig:posteriors_bent} show the resulting posterior distributions for the model parameters of the Schechter profile for the linear and the polynomial model, respectively. The distributions have qualitatively similar shapes, and the ones for CHILES are manifestly broader than the posteriors of the MIGHTEE and the combined cases, as expected from the smaller statistics. We will have a closer look to the effect of the sample size in Appendix \ref{app:sample_size}. Figs. \ref{fig:posteriors_comb}, \ref{fig:posteriors_mightee}, and \ref{fig:posteriors_chiles} display instead a comparison between the posterior distributions from the linear (blue solid) and the polynomial model (orange dashed), for the combined, MIGHTEE, and CHILES cases, respectively. Here, we clearly appreciate the systematic shift of the posteriors when passing from the linear to the polynomial model.

\section{The impact of scatter}\label{app:scatter}

\begin{figure}
    \centering
   \includegraphics[width=\columnwidth]{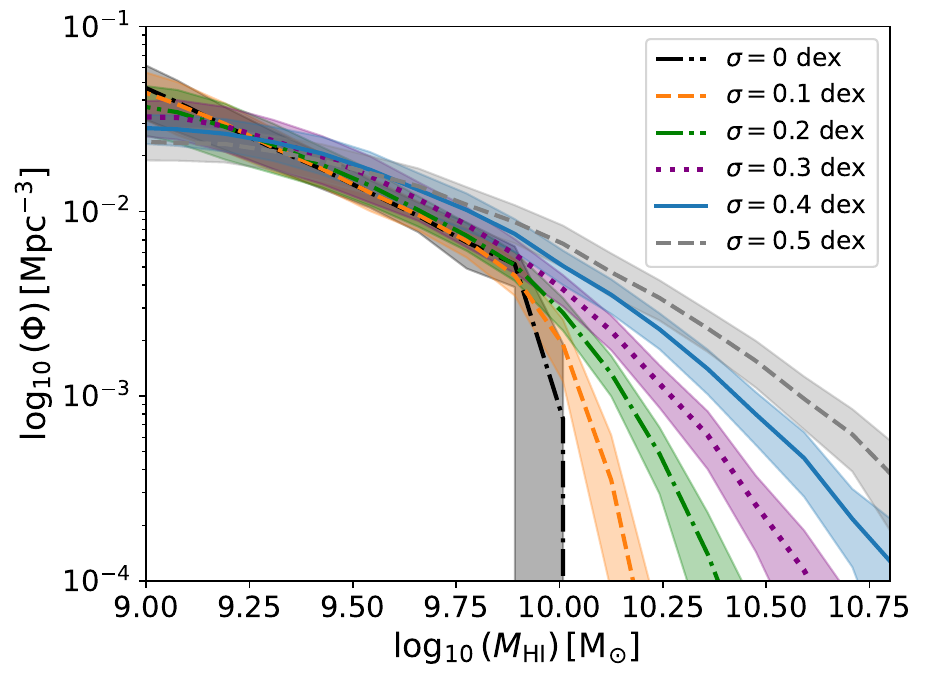}
    \caption{Results for the HIMF from the combined stacking case, as a function of the assumed scatter in the $M_{\rm HI}-M_\star$ and for the polynomial model. We show the results for $\sigma=0$ as a black dotted-dashed line, for $\sigma=0.1$ as an orange dashed line, for $\sigma=0.2$ as green dotted-dashed line, for $\sigma=0.3$ as a purple dotted line, for $\sigma=0.4$ as a solid blue line, and for $\sigma=0.5$ as a gray dashed line. Bottom: HIMF from the combined stacking assuming a polynomial and an asymmetric scattering (orange diamonds), compared to the fiducial HIMF from this work (blue squares) and to the ALFALFA HIMF at $z\sim 0$ (green dashed line).)}
    \label{fig:himf_scatter}
\end{figure}

As shown by \cite{Bera2022}, properly modelling the scatter of the scaling relation used to derive the final HI mass sample is of pivotal importance to correctly reproduce the final HIMF. In particular, \cite{Bera2022} found that assuming zero scatter on the $M_{\rm HI}-M_B$ relation used to convert from $M_B$ to $M_{\rm HI}$ severely underestimates the massive end of the ALFALFA HIMF at $z\sim 0$. Here, we build upon that experiment and apply the full forward model assuming different values for the scatter: $\sigma=\{0,0.1,0.2,0.3,0.4,0.5\}$ dex. We show the resulting HIMF in Fig. \ref{fig:himf_scatter}. One immediately sees that the HIMF is indeed very sensitive on the assumed scatter, with the high-mass end of the HIMF increasing to larger values with increasing scatter. This is due to the fact that the scatter allows to sample synthetic $M_{\rm HI}$ values from a statistical distribution at fixed $M_\star$, instead of having a deterministic estimate. In terms of HIMF, this translates in a higher probability to generate low- and high-$M_{\rm HI}$ galaxies the larger the scatter. This test shows the importance of carefully modeling the scatter and assuming an adequate value for it.

In addition, after choosing $\sigma=0.39$ as motivated by $z\sim 0$ results from the xGASS survey \citep{Catinella2018} and explained in Sect. \ref{sec:methods:mstar_to_mhi}, we also incorporate in our analysis an additional source of error coming from the intrinsic uncertainty on the true value of the scatter. In particular, for each step of our Monte Carlo approach, we add to the scatter a random error sampled from a Gaussian distribution with zero mean and standard deviation $\sigma_{\rm err}=0.05$. In this way, we forward-model this additional source of uncertainty, and propagate it throughout the whole procedure. We report In Table \ref{tab:bestfit_himf_scatter} the results of the best-fitting parameters to the HIMF for the Schechter function, with and without assuming the uncertainty on the scatter. As can be seen from the numerical results, the differences between including or not the scatter are negligible and not systematic. Specifically, the uncertainties on the best-fitting parameters remain roughly unchanged, and the median values suffer tiny random shifts, which are not statistically significant. As explained in Sect. \ref{sec:methods:mstar_to_mhi}, we hence assume as fiducial methodology the one including the error on the scatter, for completeness. 

\begin{table*}[]
    \centering
    \begin{tabular}{lcccc}
    \toprule
       Survey   & $\log_{10}(\tilde{M/{\rm M}_{\odot}})$ & $\alpha$ & $\log_{10}(\tilde{\Phi}/{\rm Mpc}^{-3})$ & 
       $\Omega_{\rm HI}\times 10^{-4}$ \\
       \midrule
       & &  w/ scatter unc. & & \\
       \midrule
       MIGHTEE &  $10.13^{+0.08}_{-0.09}$ & $-1.38^{+0.11}_{-0.10}$ & $-2.24^{+0.13}_{-0.13}$ & $8.16^{+0.59}_{-0.54}$ \\
       \addlinespace[1.5mm]
       CHILES & $10.01^{+0.16}_{-0.15}$ & $-1.31^{+0.23}_{-0.20}$ & $-2.18^{+0.21}_{-0.25}$ & $6.65^{+0.99}_{-0.69}$\\
       \addlinespace[1.5mm]
       Combined & $10.16^{+0.08}_{-0.08}$ & $-1.48^{+0.10}_{-0.09}$ & $-2.41^{+0.12}_{-0.13}$ & $7.02^{+0.59}_{-0.52}$ \\
       \midrule
        & & wo/ scatter unc. & & \\
       \midrule
       MIGHTEE &  $10.11^{+0.09}_{-0.10}$ & $-1.37^{+0.11}_{-0.10}$ & $-2.23^{+0.12}_{-0.14}$ & $8.09^{+0.60}_{-0.54}$ \\
       \addlinespace[1.5mm]
       CHILES & $10.02^{+0.16}_{-0.16}$ & $-1.33^{+0.22}_{-0.19}$ & $-2.20^{+0.21}_{-0.24}$ & $6.67^{+0.98}_{-0.80}$\\
       \addlinespace[1.5mm]
       Combined & $10.15^{+0.08}_{-0.08}$ & $-1.48^{+0.10}_{-0.09}$ & $-2.41^{+0.13}_{-0.13}$ & $6.99^{+0.61}_{-0.51}$ \\
        \bottomrule
    \end{tabular}
    \vspace{1mm}
    \caption{Best fit parameters of HIMF with and without assuming the uncertainty on the scatter}
    \label{tab:bestfit_himf_scatter}
\end{table*}


\section{The impact of the $M_{\rm HI}$ distribution asymmetry}\label{app:asymmetric}

\begin{figure}
    \centering
    \includegraphics[width=\columnwidth]{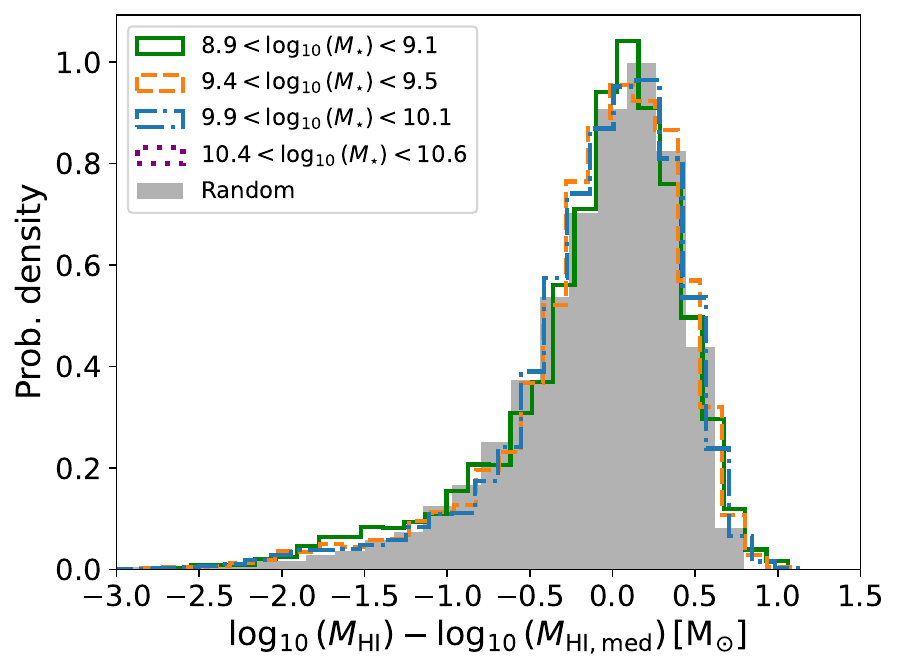}
    \includegraphics[width=\columnwidth]{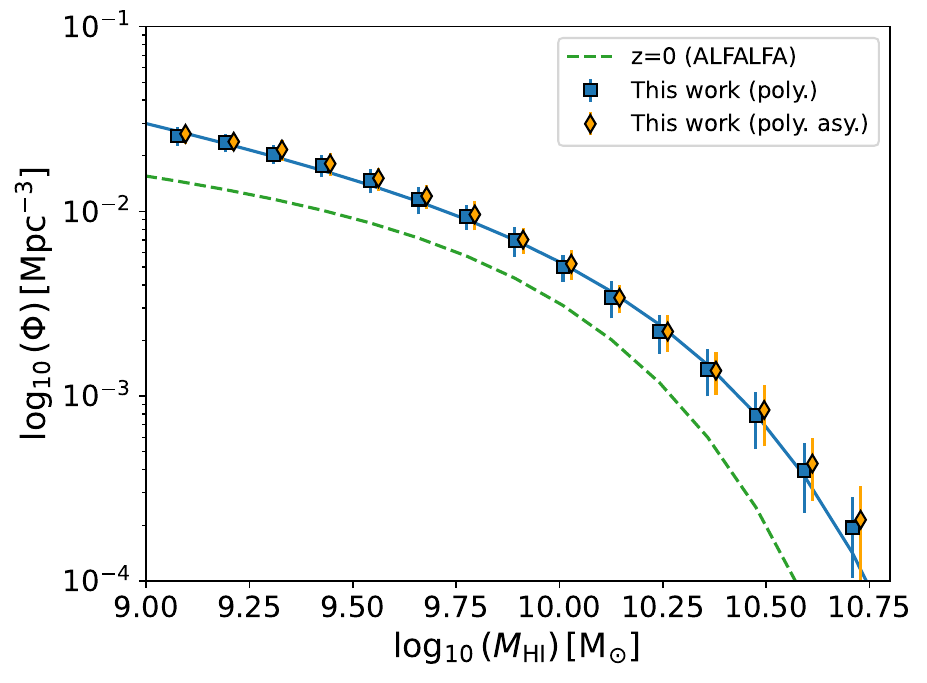}
    \caption{Results of testing the impact of an asymmetric scatter for the $M_{\rm HI}-M_\star$ relation. Top: $M_{\rm HI}$ distributions (to which we have subtracted the median) in different stellar mass bins: $8.9<\log_{10}(M_\star/{\rm M_\odot})<9.1$ (green solid), $9.4<\log_{10}(M_\star/{\rm M_\odot})<9.6$ (orange dashed), $9.9<\log_{10}(M_\star/{\rm M_\odot})<10.1$ (blue dotted-dashed), $10.4<\log_{10}(M_\star/{\rm M_\odot})<10.5$ (purple dotted). We display as a gray dashed a lognormal random distribution obtained by assuming $\sigma=0.45$, which is found to approximate well all the $M_{\rm HI}$ distributions.}
    \label{fig:asymmetry}
\end{figure}

Throughout the main analysis underlying this paper, we assumed that the $M_{\rm HI}$ distribution at fixed stellar mass is symmetric in log space, and in particular we modeled the scatter by sampling $M_{\rm HI}$ values from a Gaussian distribution. In this Appendix, we relax this assumption and test the impact of assuming an asymmetric distribution instead.

To do so, we first consider the cross-matched ALFALFA-SDSS catalog \citep{Durbala2020} and look at the $M_{\rm HI}$ distribution in different stellar mass bins. We display the results (to which we have subtracted the median, to have them centered in zero) in the top panel of Fig. \ref{fig:asymmetry}. A visual inspection of such distributions reveals that they are manifestly asymmetric, with the lower envelope featuring a larger dispersion ($\sigma\sim 0.5$ dex) than the upper envelope ($\sigma \sim 0.38$ dex). We also observe that the $M_{\rm HI}$ histograms are almost independent on stellar mass. We find that a lognormal distribution with $\sigma=0.45$ (dashed gray) represents well all the different $M_{\rm HI}$ distributions. Therefore, we repeat the main analysis replacing the Gaussian scatter with a lognormal scatter (in log space) as described above. We display the resulting HIMF (orange diamonds) in the bottom panel of Fig. \ref{fig:asymmetry}, compared to the results from assuming a Gaussian scatter (blue squares) and to the HIMF from ALFALFA at $z\sim 0$ for comparison. The outcome of this experiment highlights that the difference between assuming appropriate symmetric and asymmetric scatters is negligible in this case. Therefore, assuming a Gaussian scatter, even though simplistic, turns out to be a robust assumption in our framework.


\section{Completeness effects on the HIMF}\label{app:completeness}

\begin{figure}
    \centering
    \includegraphics[width=\columnwidth]{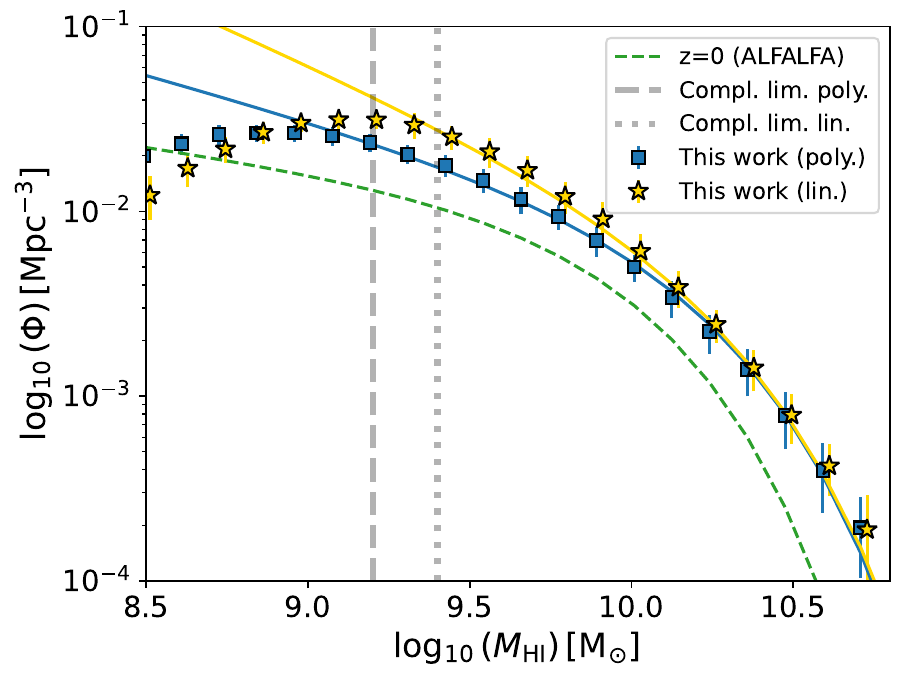}
    \caption{Completeness effects on the HIMF. We display the resulting HIMF obtained by assuming the linear (yellow stars) and the polynomial (blue squares) models for the $M_{\rm HI}-M_\star$ relation. The vertical gray dashed and dotted line stand for the conservative completeness limits $\log_{10}(M_{\rm HI}/{\rm M}_\odot)=9.2$ and $\log_{10}(M_{\rm HI}/{\rm M}_\odot)=9.4$ for the polynomial and linear cases, respectively, and represent the lower limits used for the fit of the Schechter models to the data. We show the best-fitting Schechter models obtained adopting the fitting strategy outlined above as solid yellow (linear model) and solid blue (polynomial model lines. We also display the HIMF from ALFALFA at $z\sim 0$ (green dashed) for comparison.}
    \label{fig_himf_completeness}
\end{figure}

Fig. \ref{fig_himf_completeness} shows the resulting HIMF from assuming the combined scaling relation and the linear (yellow stars) and the polynomial (blue squares) models. The HIMF presents evidence of incompleteness at $\log_{10}(M_{\rm HI}/{\rm M}_\odot)\lesssim 9$. As stated in Sect. \ref{sec:methods:fit_himf}, we conservatively fit the Schechter model only at $\log_{10}(M_{\rm HI}/{\rm M}_\odot)> 9.2$ for the polynomial model and at $\log_{10}(M_{\rm HI}/{\rm M}_\odot)> 9.4$ for the linear model. The $M_{\rm HI}$ lower limits for the fit, $\log_{10}(M_{\rm HI}/{\rm M}_\odot)=9.2$ and $\log_{10}(M_{\rm HI}/{\rm M}_\odot)=9.4$, are marked by a vertical black dashed line and a vertical black dotted line, respectively. We display the best-fitting Schechter models obtained adopting the fitting strategy outlined above as solid yellow (linear model) and solid blue (polynomial model lines.  

\section{The impact of sample size}\label{app:sample_size}

\begin{figure}
    \centering
    \includegraphics[width=\columnwidth]{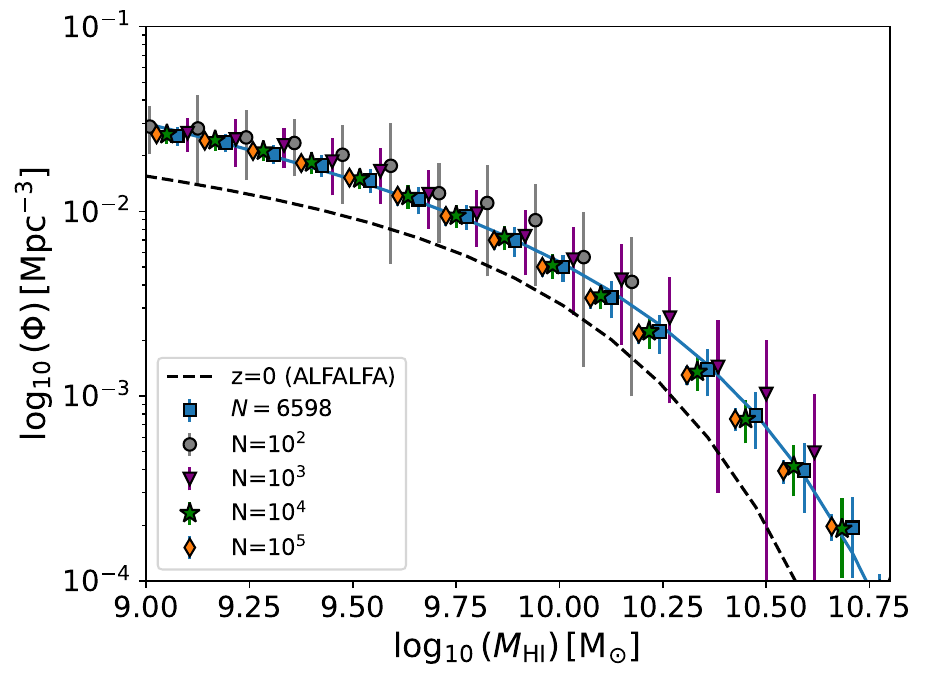}
    \caption{Results for the HIMF from the combined stacking case, as a function of the chosen sample size $N$ and assuming the polynomial model for the $M_{\rm HI}-M_\star$ relation. We show the main results from this paper ($N=6598$) as blue squares, for $N=10^2$ as gray circles, for $N=10^3$ as purple downward triangles, for $N=10^4$ as green stars, and for $N=10^5$ as orange diamonds. We also display the HIMF from ALFALFA at $z\sim 0$ as a black dashed line, for comparison. A horizontal tiny shift was applied to the data points to avoid overlaps and to ease visualization.}
    \label{fig:himf_sample_size}
\end{figure}

To assess the impact of the assumed sample size, we repeat the main procedure outline in Sect. \ref{sec:methods}, but varying the sample size. In this way, we aim at testing the impact on the HIMF of the sample size  which we matched to the one used in the analysis from \cite{Bianchetti2025}. Fig. \ref{fig:himf_sample_size} shows the results of this test. In particular, we compare the results for the combined stacking case from the `true' sample size already presented in Sect. \ref{sec:results:himf} (blue squares), to the resulting HIMF from increasingly larger samples of sizes $N=10^2$ (gray circles), $N=10^3$ (purple downward triangles), $N=10^4$ (green stars), and $N=10^5$ (orange diamonds). One sees that there are no evident systematic effects due to sample size for $N>1000$. Only in the case $N=100$ and $N=1000$ a small systematic shift of the HIMF is observed, although well within the uncertainties. In particular, for $N=100$ the HIMF is not measured for the used binning and setup at $\log_{10}(M_{\rm HI}/{\rm M}_\odot)\gtrsim 10.2$ due to an insufficient sampling. 

This test has two important implications for this work. First, it is safe to assume that for the sample sizes used in our experiment, the results are well converged, and the sample size only affects the magnitude of the uncertainty, as sought. Second, given the large uncertainties featured by the cases $N=100$ and $N=1000$, it is not surprising that current results from the literature building the HIMF with few tens or hundreds of \HI{} detections are not well converged.   

\section{Posterior distributions of the model parameters}\label{app:posteriors}

\begin{figure*}
    \centering
    \includegraphics[width=0.6\textwidth]{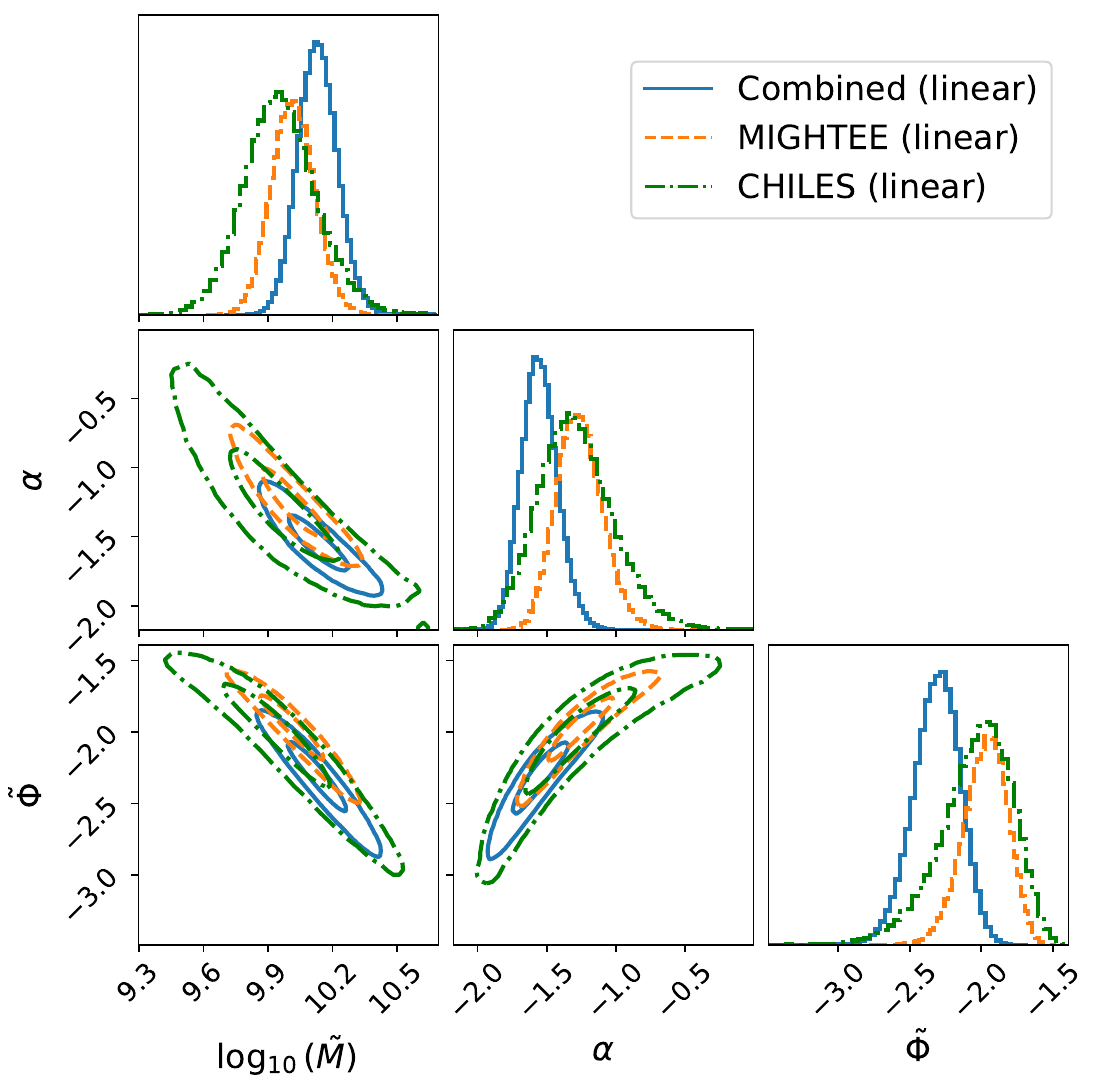}
    \caption{Posterior distributions of the best-fitting parameters of the Schechter profile to the HIMF, resulting from assuming a linear model for the $M_{\rm HI}-M_\star$ relation. The contours represent the $68\%$ and $99\%$ confidence levels.}
    \label{fig:posteriors_linear}
\end{figure*}

\begin{figure*}
    \centering
    \includegraphics[width=0.6\textwidth]{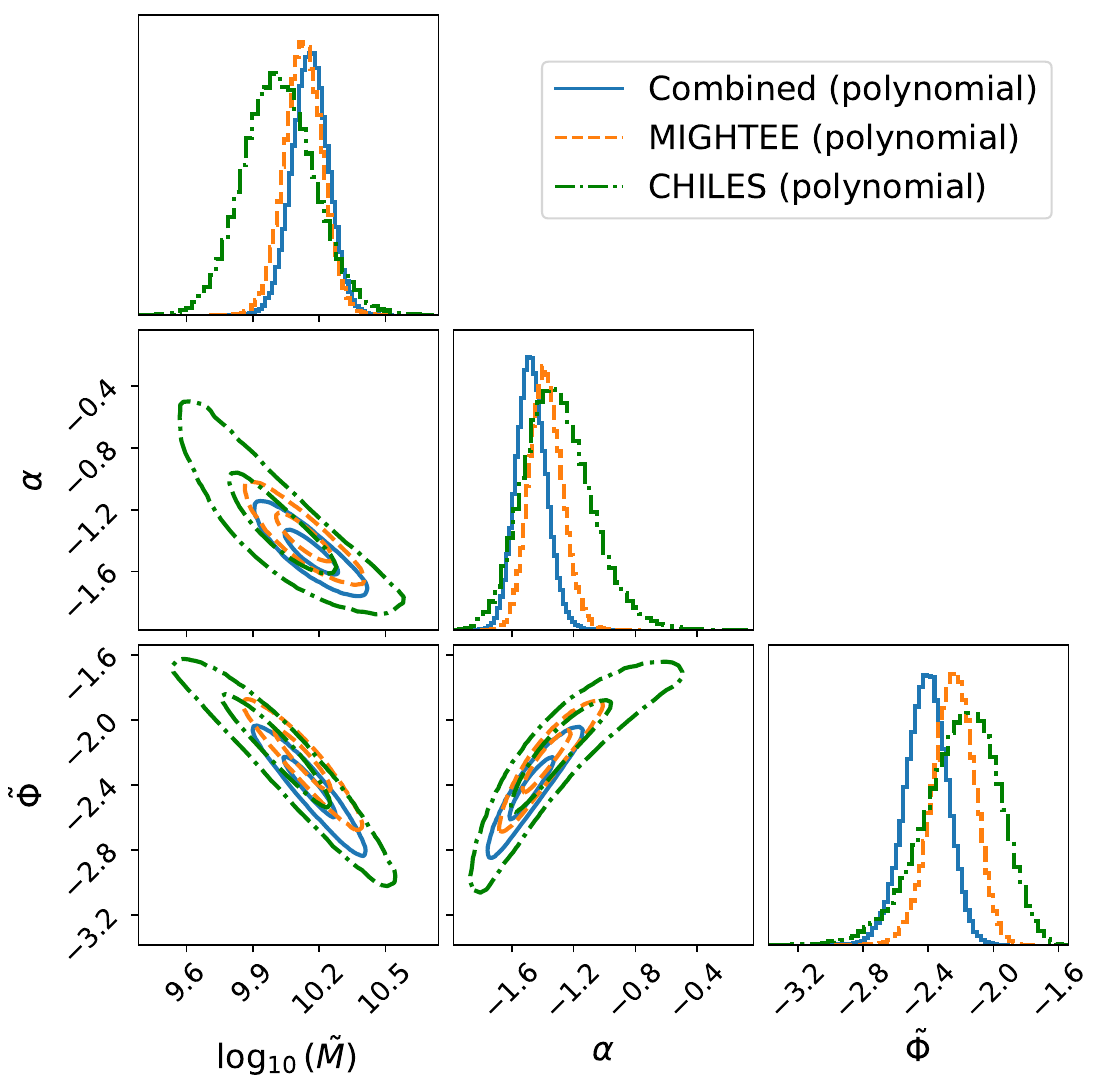}
    \caption{Posterior distributions of the best-fitting parameters of the Schechter profile to the HIMF, resulting from assuming a polynomial model for the $M_{\rm HI}-M_\star$ relation.  The contours represent the $68\%$ and $99\%$ confidence levels.}
    \label{fig:posteriors_bent}
\end{figure*}

\begin{figure*}
    \centering
    \includegraphics[width=0.6\textwidth]{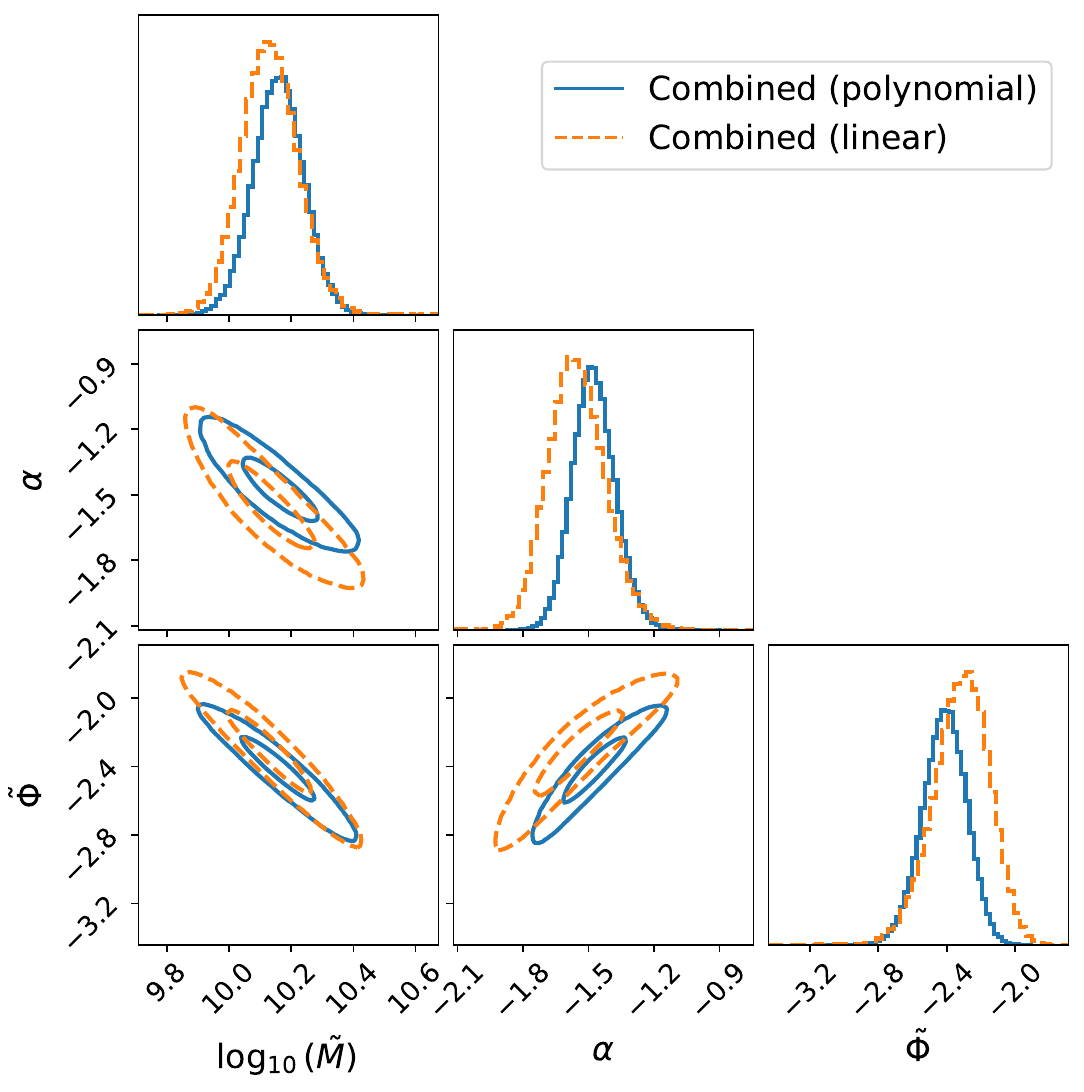}
    \caption{Posterior distributions of the best-fitting parameters of the Schechter profile to the HIMF, resulting from the combined stacking case.  The contours represent the $68\%$ and $99\%$ confidence levels.}
    \label{fig:posteriors_comb}
\end{figure*}

\begin{figure*}
    \centering
    \includegraphics[width=0.6\textwidth]{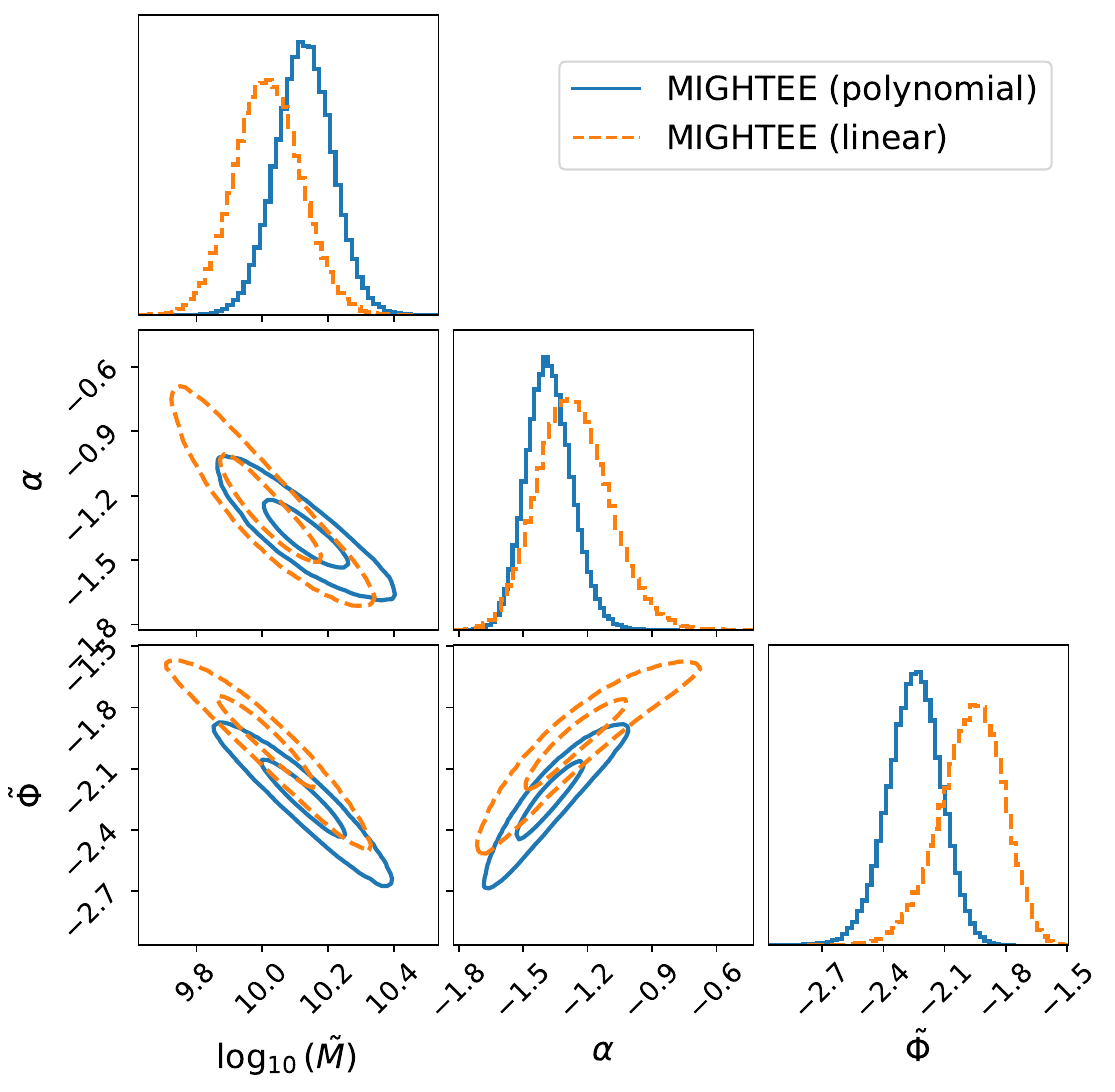}
    \caption{Posterior distributions of the best-fitting parameters of the Schechter profile to the HIMF, resulting from the MIGHTEE stacking case. The contours represent the $68\%$ and $99\%$ confidence levels.}
    \label{fig:posteriors_mightee}
\end{figure*}

\begin{figure*}
    \centering
    \includegraphics[width=0.6\textwidth]{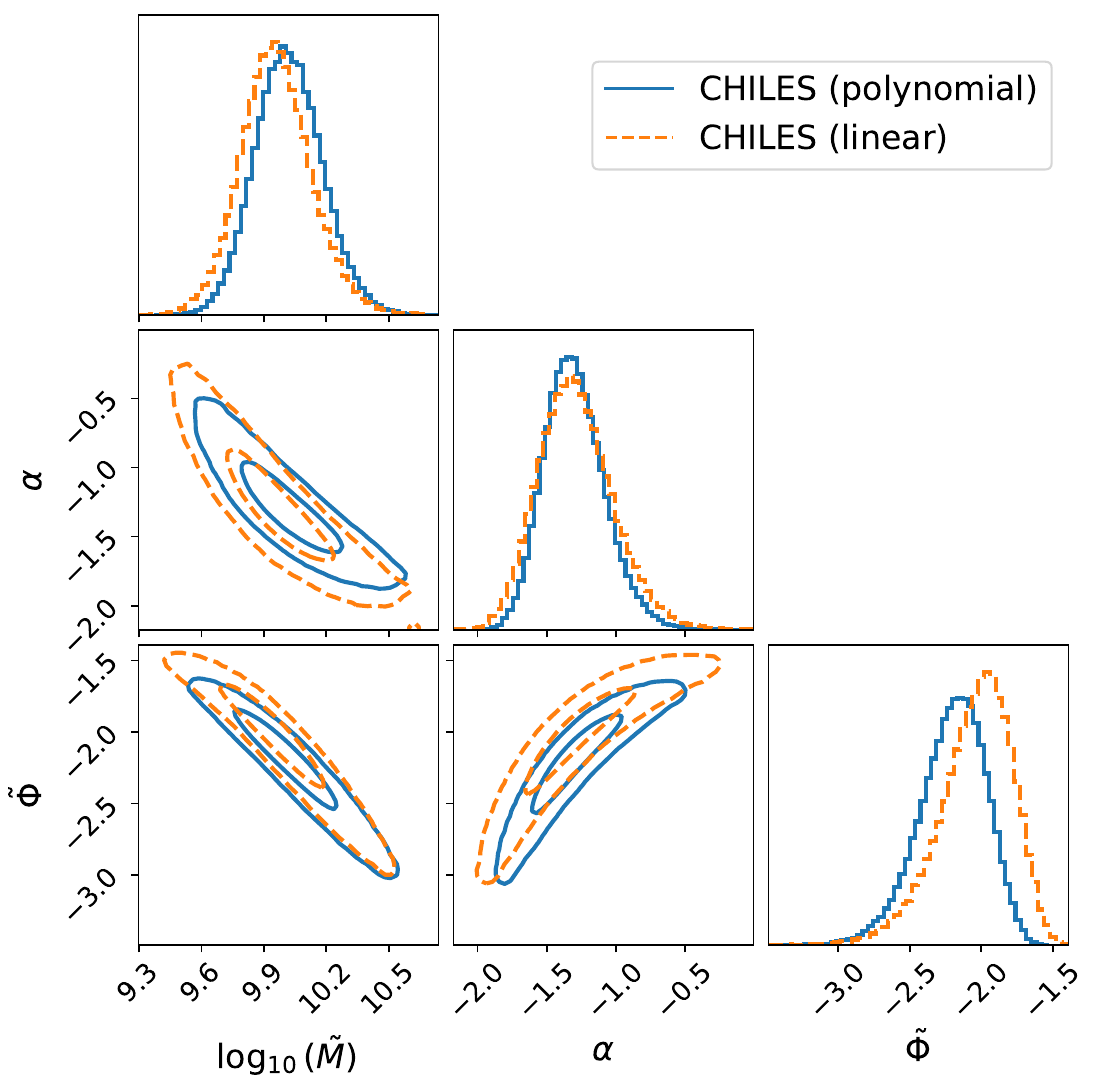}
    \caption{Posterior distributions of the best-fitting parameters of the Schechter profile to the HIMF, resulting from the CHILES stacking case.  The contours represent the $68\%$ and $99\%$ confidence levels.}
    \label{fig:posteriors_chiles}
\end{figure*}

In this Appendix, we display the posterior distributions for the different studied cases. Figs. \ref{fig:posteriors_linear} and \ref{fig:posteriors_bent} show the resulting posterior distributions for the model parameters of the Schechter profile for the linear and the polynomial model, respectively. Figs. \ref{fig:posteriors_comb}, \ref{fig:posteriors_mightee}, and \ref{fig:posteriors_chiles} display instead a comparison between the posterior distributions from the linear (blue solid) and the polynomial model (orange dashed), for the combined, MIGHTEE, and CHILES cases, respectively. Here, we clearly appreciate the systematic shift of the posteriors when passing from the linear to the polynomial model.

\end{appendix}

\end{document}